\documentclass[letter,10pt]{article}
\usepackage{graphicx,}
\usepackage{amsmath}
\usepackage[comma,authoryear,round]{natbib}
\usepackage[overload]{empheq}
\usepackage{color}
\definecolor{darkred}{rgb}{0.7,0.0,0.0}

\newcommand{\tea}{T_\mathrm{E\to A}}
\newcommand{\taf}{T_\mathrm{A\to F}}
\newcommand{\tfh}{T_\mathrm{F\to H}}
\newcommand{\thc}{T_\mathrm{H\to C}}
\newcommand{\thd}{T_\mathrm{H\to D}}
\newcommand{\tcd}{T_\mathrm{C\to D}}
\newcommand{\tao}{T_\mathrm{A\to O}}
\newcommand{\tio}{T_\mathrm{I\to O}}
\newcommand{\tfo}{T_\mathrm{F\to O}}

\newcommand{\tei}{T_\mathrm{E\to I}}
\newcommand{\tir}{T_\mathrm{I\to R}}

\newcommand{\thr}{T_\mathrm{H\to R}}
\newcommand{\tcr}{T_\mathrm{C\to R}}

\newcommand{\tih}{T_\mathrm{I\to H}}

\newcommand{\fEA}{f_\mathrm{E\to A}}
\newcommand{\fEI}{f_\mathrm{E\to I}}
\newcommand{\fAF}{f_\mathrm{A\to F}}
\newcommand{\fFH}{f_\mathrm{F\to H}}

\newcommand{\fIH}{f_\mathrm{I\to H}}

\newcommand{\fHC}{f_\mathrm{H\to C}}
\newcommand{\fCD}{f_\mathrm{C\to D}}
\newcommand{\fHD}{f_\mathrm{H\to D}}

\newcommand{\rz}{$R_0$}

\newcommand{\deps}{\emph{d\'epartements}}

\begin{document}
\begin{center}
{\LARGE
  \bf
Regional analysis of COVID-19 in France from fit of hospital data with
different evolutionary models} \\
 %% Regional analysis of COVID-19 in France from fit of
 %%  hospital data with different evolutionary models}  \\
  \bigskip
{  \large
  Gary A. Mamon}\\
  \medskip
{  \sl
  Institut d'Astrophysique de Paris (UMR 7095: CNRS \& Sorbonne
  Universit\'e)}\\
  %% \large
  %% \medskip
  %% \rm
  %% \today

\section*{Abstract}
\parbox{14cm}{
%%   The SIR %% and SEIR models are not sufficiently precise to account for the
%%   evolutionary model c
%% daily hospital data on the COVID-19 pandemic. In particular, SIR predicts too
%% sharp a decrease of the fractions of Infectious people after the start of a
%% lockdown, compared to what is observed.
The SIR evolutionary model predicts too
sharp a decrease of the fractions of  people infected with COVID-19 in France
after the start of the national 
lockdown, compared to what is observed.
 I fit the daily hospital data: arrivals in regular and critical care units,
 releases and deaths, using extended SEIR models.  
% I introduce three extensions to these models
%% : SEIHCDRO and
%% SEAFCHDRO, which both split the Infectious phase into
%% Hospitalized and Critical,
%% as well as into Asymptomatic and Feverish for the latter, and also SEAFHDRO,
%% which merges the Hospitalized and Critical phases.
%% The free parameters of these models include
%% to fit the daily arrivals in French hospitals, as well  in
%% critical care, releases and deaths.
 These
 % models
 involve
% 7, or 9
ratios of
evolutionary timescales to branching fractions,
% all
assumed uniform
throughout a country, and the basic reproduction number, \rz, before and
during the national lockdown, %% as well as a normalization 
for each region of France.
The joint-region Bayesian analysis allows precise evaluations of the
time/fraction ratios and pre-hospitalized fractions.
% (forthe first time).
%% A Markov Chain Monte Carlo sampler is run to fit these models to
%% French hospital data, consisting of daily hospital admissions and releases, per French
%% \dep,
%% i.e. general hospitalizations, critical-care
%% hospitalizations, deaths, and releases.
% The models%%  are run on % single \deps, as well as
%% on 15 regions, and on France
%% as a single zone, and
% are analyzed using Bayesian techniques. 
%% Bayesian evidence is used to infer that the more complex model is preferred,
%% even after considering its extra free parameter.
%% Analyzing the results with Bayesian techniques,
%% the models indicate that
The hospital data are well fit by the models, except
the arrivals in critical care, which decrease faster than
predicted, indicating better treatment
%. This suggests that hospitals 
% have learnt
over time.
%% to better treat COVID-19 patients without resorting to
%% critical care.
% The basic reproductive factor,
Averaged over France, the analysis yields
%% as high as
$R_0=3.4\pm0.1$ before the
% national
lockdown % of 17 March 2020, and 5 times lower during lockdown: $R_0 =
and
$0.65\pm0.04$ (90\% c.l.) during the lockdown, 
with small
regional variations.
%% 3) Contrary to the SIR model predictions of an exponential drop of the
%% fraction of Infectious people after lockdown, 
%% this fraction has reached a plateau slightly below
%% 1\% on April 15, which should remain for the first partial lifting of the
%% national lockdown on 11 May 2020.
On 11 May 2020, the Infection
Fatality Rate in France was  $4\pm1\%$ (90\% c.l.), while the
Feverish vastly outnumber the Asymptomatic, contrary to the early phases.
%6) The incubation period is below 6 days (at 95\% confidence).
Without the lockdown nor social distancing, over 2 million deaths from
COVID-19 would have occurred throughout y
France, while a lockdown that would have been enforced 10 days earlier would
have led to less than 1000 deaths.
The fraction of immunized people reached a plateau below 1\%
throughout France (3\% in Paris) by late April 2020 (95\% c.l.), suggesting
a lack of herd immunity.
%% and that a second wave of the pandemic is possible during the partial
%% lifting of the national lockdown.
The widespread availability of face masks on 11 May, when the lockdown was
partially lifted, should keep \rz\ 
below unity if at least 46\% of the population wear them outside their
home.
Otherwise, without enhanced other social distancing, a second wave is
inevitable and
cause the number of deaths to triple
between early May and October (if $R_0 = 1.2$) or even late June (if $R_0=2$).
%% The wide use of face masks should thus either decrease the pandemic or make
%% it manageable.
%% After the partial lifting of the lockdown, if \rz\  is as high as 1.5, then
%% a second wave will lead to 60 thousand deaths by mid-July and over a million
%% by October, while if \rz\ is 1.2 or lower, the pandemic is delayed with
%% deaths rising as late as  August, allowing for timely governmental response. 
}
\end{center}

%\new{Added text on \today\ in blue.}

\section{Introduction}
\label{sec:intro}

The current COVID-9 pandemic of infection and deaths caused by the SARS-2
virus is affecting all countries. Among these, France has been severely
affected, with over 25 thousand deaths.
%% , and one of the highest
%% \emph{Case to Fatality Rates} (CFRs) in the world (around
%% 20\%)
In any seriously life-threatening pandemic as COVID-19 is in France, it is
crucial to assess the parameters of the pandemic: 1) the \emph{timescales} between
different phases, e.g. the incubation timescale between the exposure to the
virus and the arrival of the first symptoms, 2) the \emph{branching fractions}, for
example, what fraction of the population that catches the virus eventually
develops symptoms, 3) the \emph{basic reproduction number}, termed \rz, which
measures the number of people infected by a single contagious person. With
these quantities, one can estimate the evolution in time of fractions of
infectious people, people in hospitals, in particular those in critical care,
and immunized people. These fractions are important as they enable national
and regional 
health authorities plan for lockdown measures to prevent too many deaths, as
well as to prevent saturation of the hospitals, especially given the 
limits in numbers of
critical care units.

The basis of epidemiological studies are based on considering different
\emph{compartments}, which I call \emph{phases}, of patient evolution. The simplest model is SIR, which
considers three  phases:
\emph{Susceptible} (not yet
infected),  \emph{Infectious}, and \emph{Removed} (the sum of
\emph{Recovered} and \emph{Dead}). The SEIR model inserts an \emph{Exposed}
phase (infected, but not yet contagious). %% These models include the
%% \emph{basic reproduction factor}, \rz, which measures how many Susceptibles
%% are  contaminated on average by a single Infectious.

Several such studies have been performed for France.
\citet{Massonnaud+20} used a SEIR model with fixed hypotheses on \rz\ to
forecast the short-term  hospital needs.
%% \citet{Fanelli&Piazza20}  first  noticed that the daily numbers of  confirmed (i.e. hospitalized) cases,
%% deaths, and released persons in China, Italy and France all followed a simple power-law recursion relation
%% $X(t+1) = \alpha\,X[t]^\beta$, where time $t$ is in days with origin for $X
%% =1$,
%% and found $\alpha = 2.173 \simeq {\rm e}$ and $\beta = 0.928$, leading to a
%% rapidly rising evolution later reach an asymptotic limit.
%% \citeauthor{Fanelli&Piazza20} also used an SIRD model, and provided best-fit
%% parameters, with simple  uncertainties from running many times a stochastic
%% minimizer.
\citet{Roques+20} also used an SIRD model, which they fit to the combination
of confirmed cases (presumably from hospital sources) and deaths, as well as
number of tested persons from another database. Their fits were restricted to
before the French lockdown. This led them to conclude
that the \emph{Infection Fatality Ratio} (IFR) is roughly 0.5\%, or perhaps
double accounting for deaths outside of hospitals. They also found
$R_0 = 3.2\pm0.1$ (95\% confidence).
The same authors \citep{Roques+20b} refined their model with later data
%% focused on the \^Ile de France region containing the capital,
and found a
lock-down value of $R_0 = 0.47\pm0.03$ (95 c.l.) as well as a fraction of
$3.7\pm1\%$ of the population of the region should be Immunized by early May.
Two studies extended SEIR models not only to handle deaths, but also split
the Infectious phase into Hospitalized and Critical \citep{Unlu+20} or with
four extra sub-phases within the Infectious phase \citep{DiDomenico+20}.
\citeauthor{Unlu+20} fit to the daily deaths data and determine
$R_0 = 3.56$ before lockdown and $R_0 = 0.74$ during lockdown (without
uncertainties).
\citeauthor{DiDomenico+20} fit the daily hospital and critical care
arrivals for the Paris (\^Ile de France) region.
Using 3 age classes, they introduce age-mixing matrices before and during
lockdown.
Their model 
assumed fixed
branching ratios.
Their analysis led to
$R_0 = 3.0\pm 0.2$ before lockdown, and
$R_0 = 0.68\pm 0.06$  during lockdown, with both values at 95\% confidence.

\citet{Salje+20} used a different approach than extended SIR or SEIR models, by  following
individual trajectories of hospitalized patients. They also split the patients by sex and age
group. \citeauthor{Salje+20} considered  the delays in reporting of the data,
assumed that the time from hospital arrival to death for those who do not
survive is a mixture model of an exponential distribution for rapid deaths
and a log-normal for slower ones. Their model involves a contact matrix
between people of different age groups, which was known before the pandemic,
and which they modeled after the lockdown.
They obtained \rz\ factors before lockdown of
$R_0 = 3.3\pm 0.13$ (national model) and $R_0 = 3.4\pm 0.1$ (regional model),
as well as $R_0 = 0.52\pm 0.03$  during lockdown (both models), with
95\% confidence uncertainties. They also estimate the IFR at 0.7\% and that
roughly 4\% of the population is Immunized on 11 May 2020.
%% In their regional model, these reproduction factors were
%% before lockdown, and
%% $R_0 = 0.52\pm 0.02$  during lockdown, again with both values at 95\% confidence.

These previous studies, while very useful, all failed to consider
1) different timescales for different branches of phase evolution (except
\citealp{DiDomenico+20} who assumed values for them);
2) geographic
variations in the \rz\ factor.
Indeed, on one hand, there is no reason why two branches should have equal
mean durations, and on the other, one expects that \rz\ will depend on the
local history of the pandemic as well as on the density of the zone
(i.e. higher \rz\ in the capital).

%% The study of \citeauthor{Salje+20}, while highly impressive, has several
%% drawbacks.
%% 1) They assume that \rz\ is uniform throughout France, while, before
%% lockdown,
%% it should be
%% much higher in dense zones such as the capital.
%% 2) Their zonal analysis is based on regions, and not \emph{d\'epartements}.
%% 3) They do not make predictions for the partial lifting of lockdown.

In the present study, I adapt the simple SIR and SEIR epidemiological models
to account for the daily hospital data for COVID-19 patients: arrivals
in general care, arrivals in critical care, deaths, and releases. My models
differ from previous ones, by considering different
\rz\ values per geographic zone, while assuming that the evolutionary
timescales and branching fractions are independent of the geographic zone,
i.e. that the population is genetically homogeneous.
I can then estimate how many people would have died in France without the
lockdown, how many deaths could have been prevented had the lockdown been
enforced 10 days earlier (two days after March 5, when the situation of the
pandemic in France was clearly in exponential growth with a full region
--- around Mulhouse --- infected).
I can also estimate regional disparities in pre-lockdown and lockdown values of
\rz, as well as the CFRs and Infectious-to-Fatality rates (IFRs) both
nationally and per geographic zone.
Contrary to previous models, my model does not assume that the timescales for
evolution on one or another of two branches are the same. But while my model
is less restrictive on the parameters, we shall see that it provides fairly
low uncertainties on the \rz\ factors and on the evolution in time even in
the future.

This article is organized as follows:
In Sect.~\ref{sec:Epimodels}, I present the evolutionary models, extended to
include the hospital data.
I describe the hospital data in Sect.~\ref{sec:data}.
In Sect.~\ref{sec:analysis}, I describe the analysis methods.
The results are given in Sect.~\ref{sec:results}, for the different models
run on global national data, as well as the country split into 15 regions.
%% and
%% then into 96 \deps.

\section{Evolutionary models}
\label{sec:Epimodels}

\subsection{Review of the SIR model}
\begin{figure}[ht]
  \centering
  \includegraphics[width=0.6\hsize]{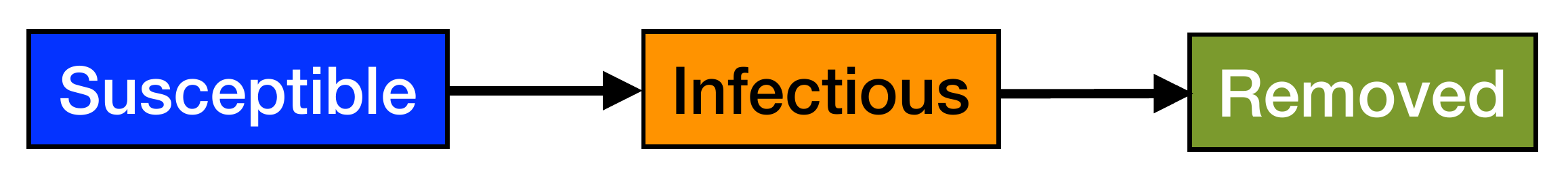}
  \caption{Illustration of the SIR model}
  \label{fig:SIR}
\end{figure}

The SIR model, illustrated in Figure~\ref{fig:SIR}, is the simplest epidemiological model.
One can write differential equations for the temporal variations of the
\emph{fractions} of people in different populations.
\begin{subequations}
  \begin{align}[left = \empheqlbrace\,]
{{\rm d}S \over{\rm d}t} &= - \beta \,S\,I  \ , \label{SdotSIR} \\
{{\rm d}I \over{\rm d}t} &= \beta\, S\,I - \gamma \, I \ , \label{IdotSIR}\\
{{\rm d}R \over{\rm d}t} &= \gamma \, I \ . \label{RdotSIR}
\end{align}
\end{subequations} 
Equation~(\ref{SdotSIR}) states that the Susceptibles are converted
into Infectious when they run into an Infectious, where $\beta$ is the
transmission rate, i.e. the (average) number of contacts between
Susceptibles and Infectious that lead to the infection of the Susceptible,
per Susceptible and per Infectious. 
Equation~(\ref{IdotSIR}) converts the loss of Susceptibles into a gain of
Infectious, but also has a loss term to account for transition to the Removed
category, either by Recovery or by Death.  Here $\gamma$ is the removal
  rate, so that $1/\gamma$ is the typical period (e.g. in days) that a person
remains Infectious.
Finally, equation~(\ref{RdotSIR}) expresses the loss of Infectious as a gain
for the Removed.

\begin{figure}[ht]
  \centering
  \includegraphics[width=0.55\hsize]{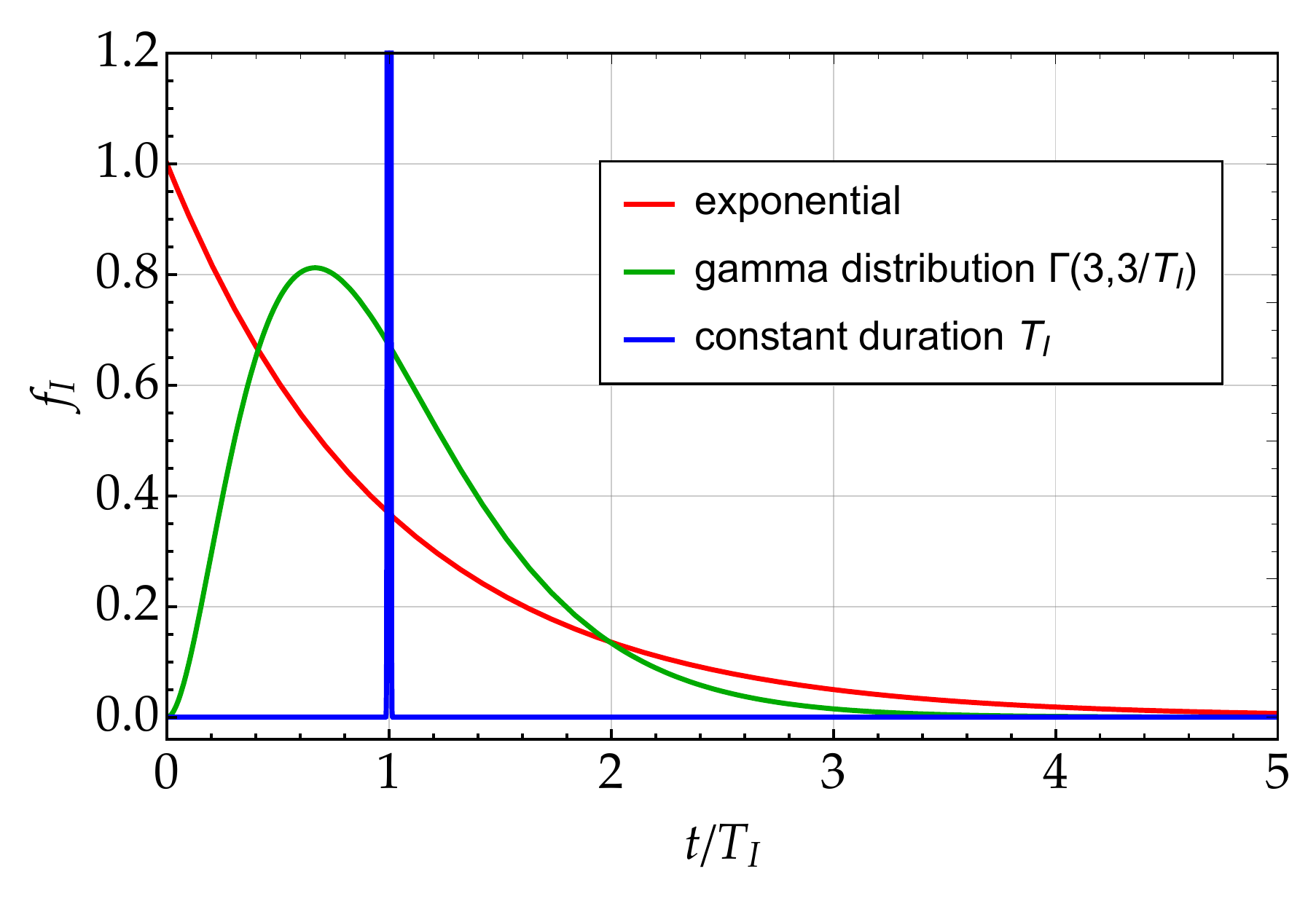}
  \caption{Different models for the distribution of the durations of the
    Infectious phase. The SIR model uses the exponential model. }
  \label{fig:fIoft}
\end{figure}

The $\gamma$ factor in equations~(\ref{IdotSIR}) and (\ref{RdotSIR})
  assumes that the time that a given Infectious becomes Removed is random,
  so one can only assert that the overall rate of transformation of
  Infectious to Recovered is proportional to the overall
  number of Infectious.
  In the absence of new infections, i.e. if $\beta=0$, the solution to
  equation~(\ref{IdotSIR}) is that the fraction of Infectious
  decreases exponentially in time: $I(t) =
  I(t_1)\,\exp\left[-\gamma(t-t_1)\right]$. This is the same equation as that of
  \emph{radioactive decay} (which is experimentally verified).
  The time $t$ is unbounded. Therefore, instead of
  a popular conception that duration of infection is constant, it follows
  instead an exponential distribution in the SIR model.
%  We will return to this issue in Sect.~\ref{sec:endconfine}.
This explains why a 15-day fully-efficient lockdown is insufficient to
eradicate a virus.
This is illustrated in Figure~\ref{fig:fIoft}, which shows the distribution
of infectious times for 3 models: exponential (as in SIR), single-valued (as
many people na\"{\i}vely believe) and a gamma distribution that corresponds
to the convolution of two identical exponential distributions.
At 3 times the mean Infectious duration, the fraction of remaining Infectious
is 5\% in the exponential model and 0.6\% in the gamma model. Quarantine
times are set to ensure that nearly all Infectious recover, but one should
remember that in models where the distribution of the infectious time is
exponential or convolutions of exponentials,
there is always a small fraction of Infectious that remain so at the end of
the quarantine.

Insight is obtained by considering the  \emph{basic reproduction number},
which
is defined as the ratio of the
transmission to removal rates:
\begin{equation}
  R_0 = {\beta \over \gamma} \ .
  \label{Rzero}
\end{equation}
Equation~(\ref{Rzero}) can also be re-written as $R_0$ being the ratio of the
infectious period to the time between infectious contacts of an Infectious with 
Susceptible people.
The differential equations can then be written
\begin{subequations}
  \begin{align}[left = \empheqlbrace\,]
\dot S &= - R_0 \,S\,{I\over \tir}  \ , \label{SdotSIR2} \\
\dot I &= \left( R_0\,S - 1\right)\, {I\over \tir} \ , \label{IdotSIR2}\\
\dot R &= {I \over \tir} \ . \label{RdotSIR2}
\end{align}
\end{subequations}

The fraction of Infectious grows as long as $R_0 > 1/S$. At the beginning of
the pandemic, $S \simeq 1$, and the criterion for growth is $R_0>1$. If the
fraction of Susceptibles is appreciably decreased, the condition for
continued growth of the pandemic is $R_0 > 1/S > 1$.

% ** Shortcomings of the SIR model... ***

\begin{figure}[ht]
  \centering
  \includegraphics[width=0.45\hsize]{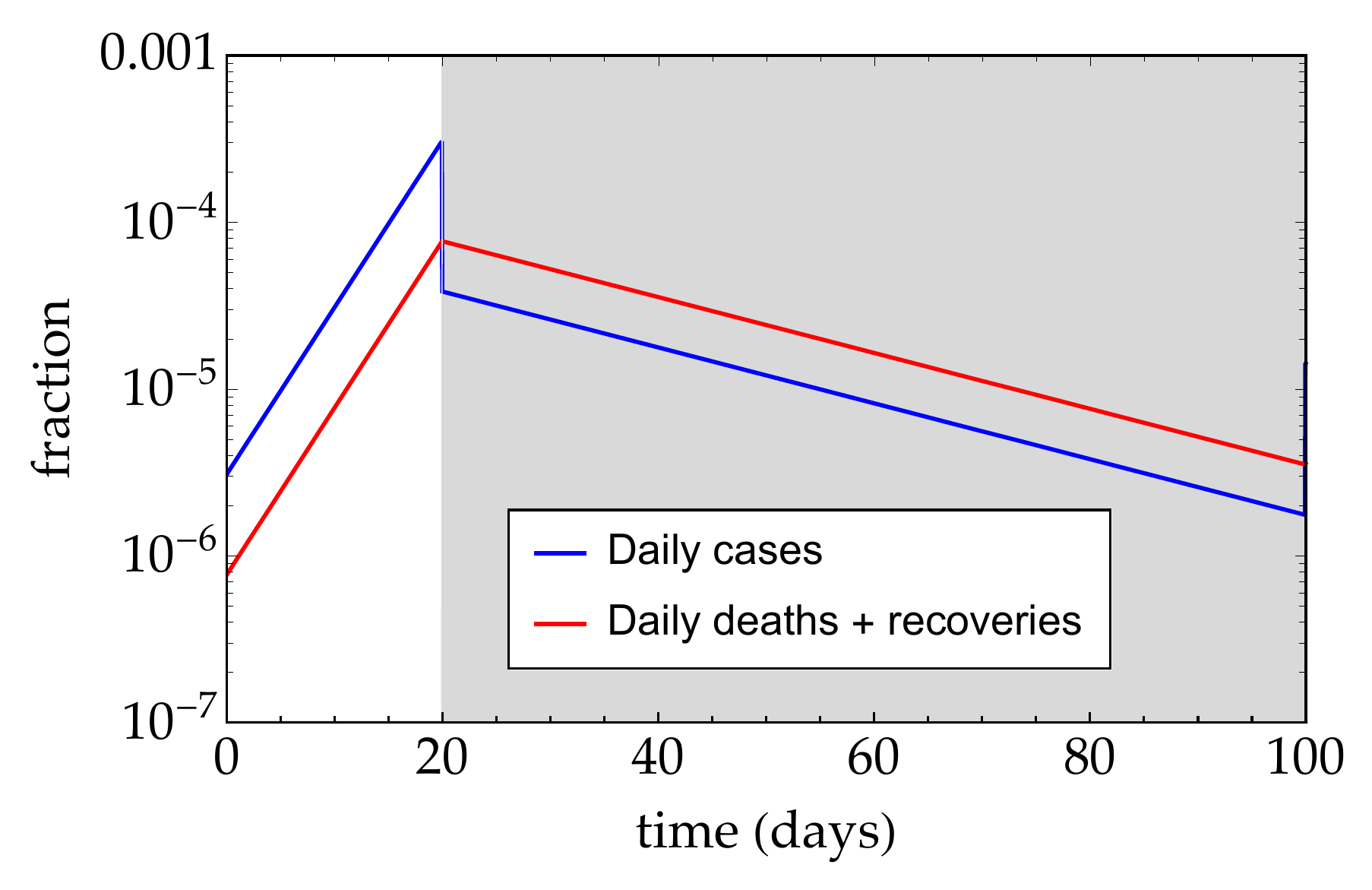}
  \includegraphics[width=0.54\hsize]{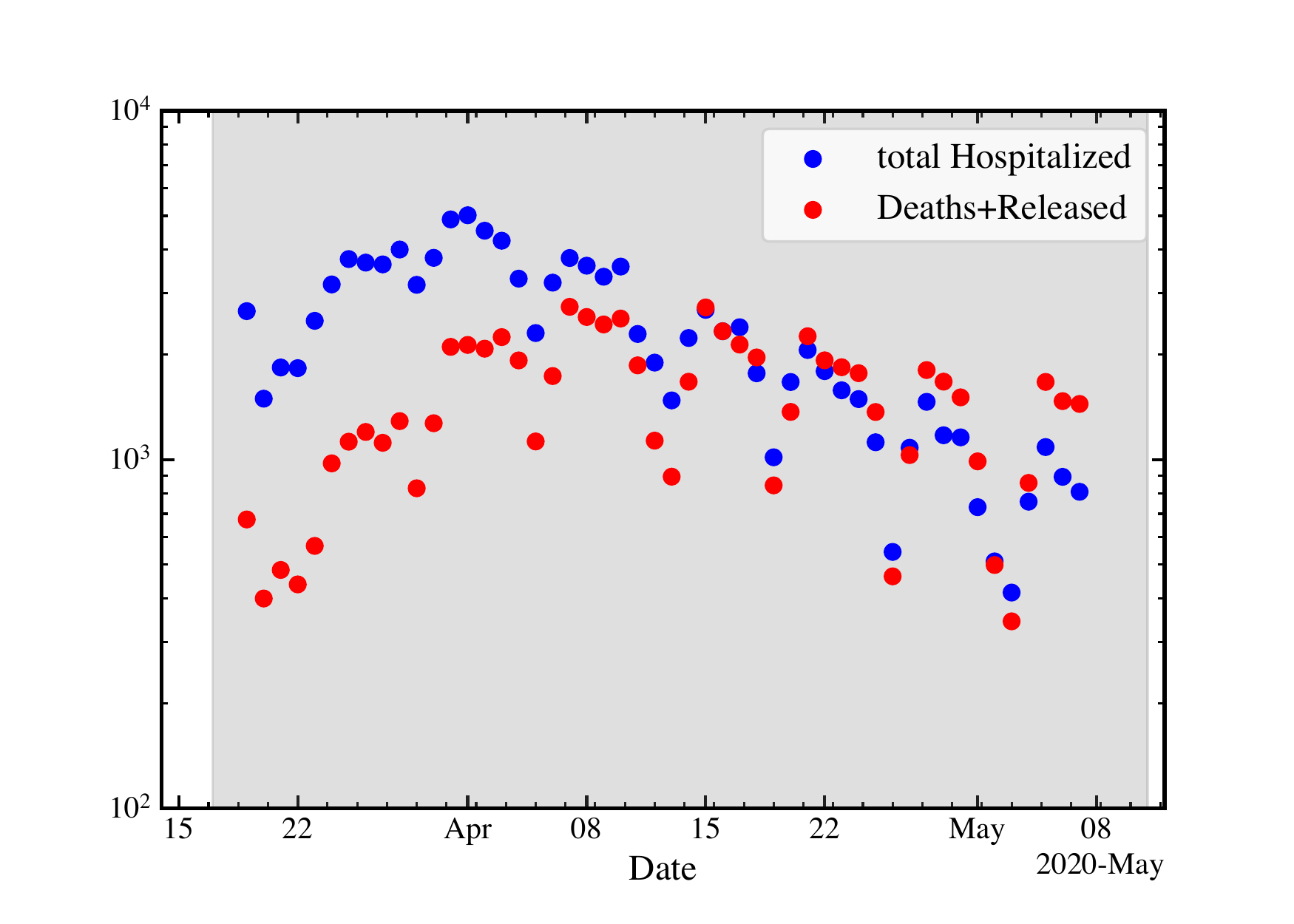} 
  \caption{
{\bf Left}:    Evolution of daily cases, $-\dot S$, (\emph{blue}) and deaths plus recovered,
$\dot R$, (\emph{red}) in the SIR model, for $R_0 = 4$ before lockdown and $R_0 = 0.5$
during lockdown, with $\tir = 5\,\rm days$.
{\bf Right}: Observed evolution of daily cases for France for
hospitalizations (\emph{blue}) and deaths plus releases (\emph{red}).
In both panels, the lockdown is shown in the
(\emph{shaded grey region}).
The decrease in observed cases and removed (deaths plus released)
respectively occurred 14 and 25 days after the start of the lockdown,
contrary to SIR prediction of immediate effects.
  }
\label{fig:DailySIR}
\end{figure}

A numerical integration of the daily cases ($-\dot S$) and daily deaths and
recoveries ($\dot R$) predicted from the SIR model in a context of a lockdown phase is
illustrated in Figure~\ref{fig:DailySIR}.  In the left panel, one sees that the SIR model
predicts an immediate sharp decrease in the daily Removed (deaths and
recoveries) with a discontinuous drop  in new cases at the start of lockdown
and a parallel but lower number of new cases relative to new deaths.
The hospital data for all of France (right panel)
indicates instead that both the new cases and daily Removed keep increasing
after the start of the lockdown for roughly 14 and 25 days, respectively.
The subsequent decreases are slower than in the SIR predictions. We thus
begin our modeling with a SEIR model, extended to include hospital data.

\subsection{SEIHCDRO model}

\begin{figure}[ht]
  \centering
    \includegraphics[width=0.8\hsize]{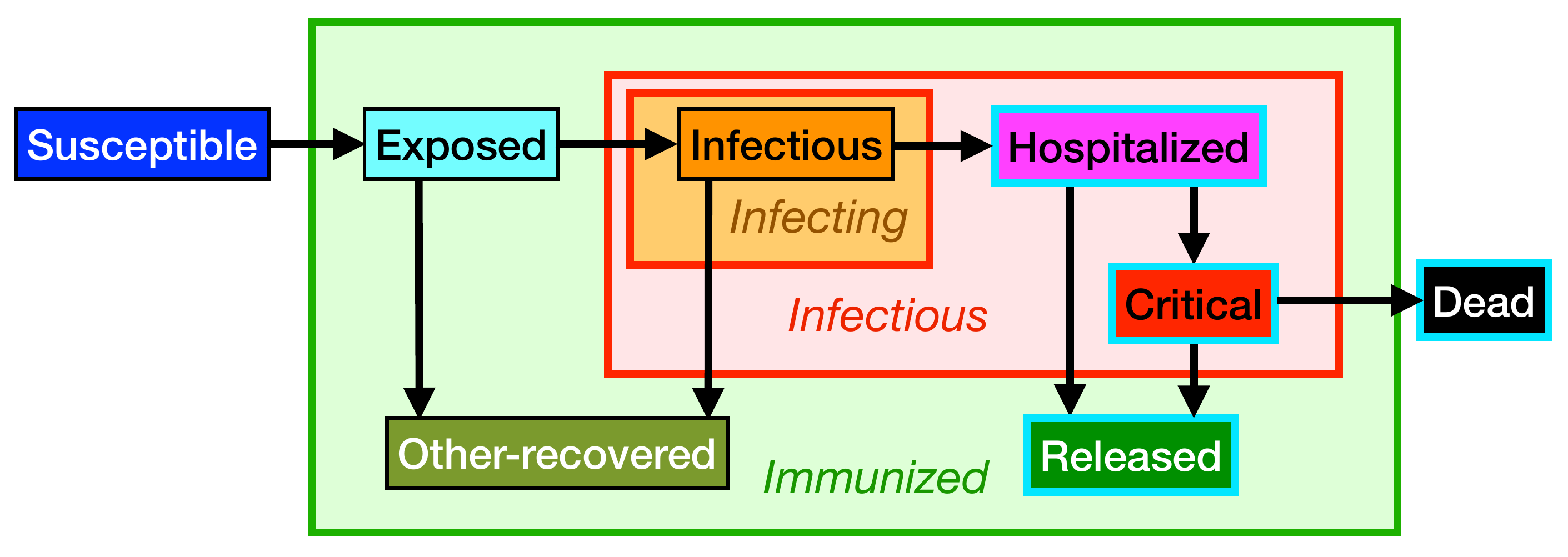}
  \caption{
    Illustration of the SEIHCDRO model.
    % See Fig.~\ref{fig:SIHCDRO} for details.
    The Infectious,
  Hospitalized and Critical phases are deemed Infectious, but only the
  Infectious phase people effectively infect the
  Susceptibles.
  Hospital data is available for the 4 phases in \emph{cyan}-surrounded boxes.
 Also shown in the large \emph{red} and \emph{green} boxes are the
  \emph{Infectious} and \emph{Immunized} super-phases.
%  All phases except Susceptible and Dead are assumed Immunized from a new infection. 
   }
\label{fig:SEIHCDRO}
\end{figure}

%% It is advisable to incorporate an additional, non-contagious, incubation
%% phase.
We extend the SEIR model (where E is for Exposed but not contagious) with
the SEIHCDRO model, which includes the 7 following phases:
%\begin{itemize}
\begin{description}
  \itemsep=0pt
  \item [Susceptibles (S)] People who may catch the virus infection, without
    being immune to it.
  \item [Infectious (I)] People who are in an
    infectious stage, but not hospitalized
  \item [Hospitalized (H)] People who are treated for COVID-19 in hospitals,
    but are not in the critical care.
  \item [Critical (C)] People who are in a critical care at a hospital.
  \item [Dead (D)] People who died of COVID-19 at a hospital.
  \item [Released (R)] People who have been released from a hospital for
    COVID-19 treatment.
  \item [Other-Recovered (O)] People who have recovered from the virus, without having
    passed through a hospital.
\end{description}
%% \begin{description}
%%   \item [Exposed (E)] People who have been exposed to the virus without having
%%     become infectious.
%% \end{description}
The SEIHCDRO model, illustrated in Figure~\ref{fig:SEIHCDRO}, includes
branching fractions for transitions one phase to two possible others.
Only a fraction of Exposed become Infectious (the remaining are similar to
the Other-Recovered). Only a fraction of Infectious are Hospitalized, while
the remaining eventually recover and join the Other-Recovered phase.
Only a fraction of Hospitalized go to Critical care, while the remaining
eventually are Released. Only a fraction of Critical people die, while the
remaining are Released.
The Other-Recovered are kept separate from the Released, because we only have
data on the latter.

In the SEIHCDRO model, the infection of hospital workers by people in the Hospitalized
and Critical phases is assumed to be negligible in comparison with the infection of
the general population by the Infectious phase people.
Indeed, %% we shall see that the Hospitalized and Critical phase populations are
%% an order of magnitude lower than that if the Infectious phase. Moreover,
Infectious people, who are often not aware of the contagiousness, meet many
more Susceptibles than do Hospitalized or Critical phase people. %% Both effects
%% combine to safely neglecting the contamination of Susceptibles by
%% Hospitalized and Critical people.
%% In the models
%% introduced later in this article, only the first  ones that follow all assume that one the first
%% infectious phase (here the Infectious one) contaminates the Susceptibles.

The equations of the SEIHCDRO model are
\begin{subequations}
  \begin{align}[left = \empheqlbrace\,]
    \dot S &= - R_0 \,S\,{I \over \overline T_{\rm I\to}}
    = - R_0\,S\,\left[{\fIH\over \tih} + {(1-\fIH)\over \tio}\right]\,I
    \ , \label{SdotSEIHCDRO} \\
    \dot E &= R_0 \,S\,{I \over \overline T_{\rm I\to}} - \fEI\,{E \over
      \tei} 
    = R_0 \,S\,\left[{\fIH\over \tih} + {(1-\fIH)\over \tio}\right]\,I -  \fEI\,{E \over
      \tei}
    \ ,\label{EdotSEIHCDRO} \\
    \dot I & =  \fEI\,{E \over \tei} - {I \over \overline T_{\rm I\to}} =
\fEI\,{E \over \tei} - \left[{\fIH\over \tih} + {(1-\fIH)\over \tio}\right]\,I
    \ , \label{IdotSEIHCDRO} \\
\dot H & =  \textcolor{darkred}{\fIH\,{I \over \tih}} - \fHC\,{H \over \thc} -
(1-\fHC)\,{H \over \thr} \ , \label{HdotSEIHCDRO} \\
\dot C & =  \textcolor{darkred}{\fHC\,{H \over \thc}} - \fCD\,{C \over \tcd} -
(1-\fCD)\,{C \over \tcr} \ , \label{CdotSEIHCDRO} \\
\dot D & =  \textcolor{darkred}{\fCD\,{C \over \tcd}} \ ,\label{DdotSEIHCDRO} \\ 
\dot R &= \textcolor{darkred}{(1-\fHC)\,{H\over \thr}} +
\textcolor{darkred}{(1-\fCD)\,{C\over \tcr}}
%- (1-\fRI)\,{R\over \trn}
\ , \label{RdotSEIHCDRO} \\
  \dot O &= (1-\fIH)\,{I\over \tio} 
\ . \label{OdotSEIHCDRO}
\end{align}
\end{subequations}
In
% \textcolor{red}{red}
red are highlighted the terms contributing to the hospital data.
Since $S+E+I+H+C+D+R+O=1$, one can omit equation~(\ref{OdotSEIHCDRO}) and
determine the Other-Recovered fraction, using $O = 1 -  (S+E+I+H+C+D+R)$.
The differential equations are written assuming that the timescales are all
different even for evolution of one phase into two possible future phases.
One notes that the independent parameters are the \rz\ factor as well as
7 ratios of timescales to
branching fractions, i.e.
$\tei/\fEI$,
$\tih/\fIH$,
$\thc/\fHC$,
$\tcd/\fCD$,
$\thr/(1-\fHC)$,
$\tcr/(1-\fCD)$,
and
$\tio/(1-\fIH)$.
Thus, the timescales and conjugate branching fractions are degenerate, but a
constraint on the ratio provides an upper limit to the timescale.

\subsection{The SEAFHCDRO model}

\begin{figure}[ht]
  \centering
    \includegraphics[width=0.9\hsize]{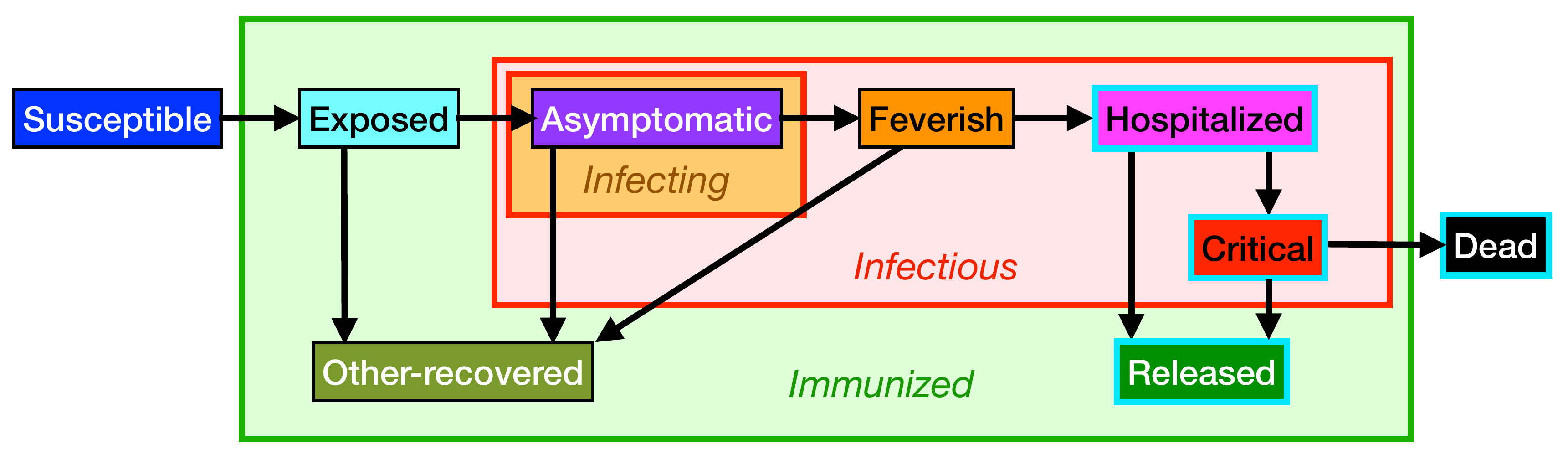}
  \caption{
Illustration of the SEAFHCDRO model, with same comments as in
Fig.~\ref{fig:SEIHCDRO}. Only the Asymptomatics effectively infect the Susceptibles.
  }
\label{fig:SEAFHCDRO}
\end{figure}

The SEIFHCDRO model can be generalized  by splitting the Infectious
compartment into one where people effectively contaminate the Susceptibles,
and another where people are too ill to go outside and contaminate
Susceptibles.
These new compartments are called the
\emph{Asymptomatic} and \emph{Feverish} phases (see
Fig.~\ref{fig:SEAFHCDRO}). 
%% We generalize the SEIFHCDRO model by splitting the Infectious stage into 
%%    \emph{Asymptomatic} and \emph{Feverish} phases, where the Feverish people do not
%% contaminate the Susceptibles, because they don't leave their dwelling (see
%% Fig.~\ref{fig:SEAFHCDRO}). 
%% \begin{description}
%%   \itemsep=0pt
%%   \item [Susceptibles (S)] People who may catch the virus infection, without
%%     being immune to it.
%%   \item [Exposed (E)] People who have been exposed to the virus without having
%%     become infectious.
%%   \item [Asymptomatic (A)] People who are in an
%%     infectious stage, but with no symptoms.
%%   \item [Feverish (F)] People who show COVID-19 symptoms, but have not been
%%     hospitalized.
%%   \item [Hospitalized (H)] People who are treated for COVID-19 in hospitals,
%%     but are not in the critical care.
%%   \item [Critical (C)] People who are in a critical care at a hospital.
%%   \item [Dead (D)] People who died of COVID-19 at a hospital.
%%   \item [Released (R)] People who have been released from a hospital for
%%     COVID-19 treatment.
%%   \item [Others (O)] People who have recovered from the virus, without having
%%     passed through a hospital.
%% \end{description}

The equations of the SEAFHCDRO model are
\begin{subequations}
  \begin{align}[left = \empheqlbrace\,]
    \dot S &= - R_0 \,S\,{A \over \overline T_{\rm A\to}}
    = - R_0\,S\,\left[{\fAF\over \taf} + {(1-\fAF)\over \tao}\right]\,A
    \ , \label{SdotSEAFHCDRO} \\
    \dot E &= R_0 \,S\,{A \over \overline T_{\rm A\to}} - \fEA\,{E \over
      \tea}
    = R_0 \,S\,\left[{\fAF\over \taf} + {(1-\fAF)\over \tao}\right]\,A -  \fEA\,{E \over
      \tea}
    \ ,\label{EdotSEAFHCDRO} \\
    \dot A & =  \fEA\,{E \over \tea} - {A \over \overline T_{\rm A\to}} =
\fEA\,{E \over \tea} - \left[{\fAF\over \taf} + {(1-\fAF)\over
    \tao}\right]\,A
\ , \label{AdotSEAFHCDRO} \\
\dot F &= \fAF\,{A\over \taf} - \fFH\,{F\over \tfh} - (1-\fFH){F \over \tfo}\
\ , \label{FdotSEAFHCDRO} \\
\dot H & =  \textcolor{darkred}{\fFH\,{F \over \tfh}} - \fHC\,{H \over \thc} -
(1-\fHC)\,{H \over \thr} \ , \label{HdotSEAFHCDRO} \\
\dot C & =  \textcolor{darkred}{\fHC\,{H \over \thc}} - \fCD\,{C \over \tcd} -
(1-\fCD)\,{C \over \tcr} \ , \label{CdotSEAFHCDRO} \\
\dot D & =  \textcolor{darkred}{\fCD\,{C \over \tcd}} \ ,\label{DdotSEAFHCDRO} \\ 
\dot R &= \textcolor{darkred}{(1-\fHC)\,{H\over \thr}} +
\textcolor{darkred}{(1-\fCD)\,{C\over \tcr}}
%- (1-\fRI)\,{R\over \trn}
\ , \label{RdotSEAFHCDRO} \\
  \dot O &= (1-\fAF)\,{I\over \tao} + (1-\fFH){F\over \tfo}
\ . \label{OdotSEAFHCDRO}
\end{align}
\end{subequations}

\subsection{The SEAFHDRO model}

\begin{figure}[ht]
  \centering
    \includegraphics[width=\hsize]{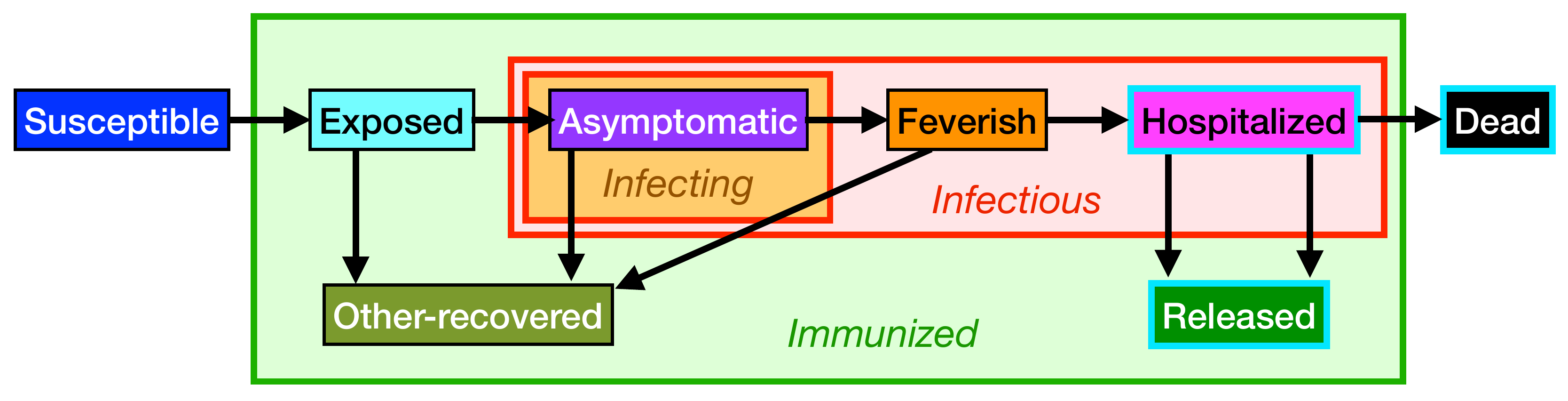}
  \caption{Illustration of the SEAFHDRO model, with same comments as Fig.~\ref{fig:SEAFHCDRO}}
\label{fig:SEAFHDRO}
\end{figure}

We shall see below that neither the SEIFHCDRO nor the SEAFHCDRO models are
able to fit the evolution of the Critical population.
Our final SEAFHDRO model (Fig.~\ref{fig:SEAFHDRO}) therefore eliminates the
Critical phase, by merging it with the Hospitalized phase.
Its  equations are
\begin{subequations}
  \begin{align}[left = \empheqlbrace\,]
    \dot S &= - R_0 \,S\,{A \over \overline T_{\rm A\to}}
    = - R_0\,S\,\left[{\fAF\over \taf} + {(1-\fAF)\over \tao}\right]\,A
    \ , \label{SdotSEAFHDRO} \\
    \dot E &= R_0 \,S\,{A \over \overline T_{\rm A\to}} - \fEA\,{E \over
      \tea}
    = R_0 \,S\,\left[{\fAF\over \taf} + {(1-\fAF)\over \tao}\right]\,A -  \fEA\,{E \over
      \tea}
    \ ,\label{EdotSEAFHDRO} \\
    \dot A & =  \fEA\,{E \over \tea} - {A \over \overline T_{\rm A\to}} =
\fEA\,{E \over \tea} - \left[{\fAF\over \taf} + {(1-\fAF)\over
    \tao}\right]\,A
\ , \label{AdotSEAFHDRO} \\
\dot F &= \fAF\,{A\over \taf} - \fFH\,{F\over \tfh} - (1-\fFH){F \over \tfo}\
\ , \label{FdotSEAFHDRO} \\
\dot H & =  \textcolor{darkred}{\fFH\,{F \over \tfh}} - \fHD\,{H \over \thd} -
(1-\fHD)\,{H \over \thr} \ , \label{HdotSEAFHDRO} \\
\dot D & =  \textcolor{darkred}{\fHD\,{H \over \thd}} \ ,\label{DdotSEAFHDRO} \\ 
\dot R &= \textcolor{darkred}{(1-\fHD)\,{H\over \thr}} 
%- (1-\fRI)\,{R\over \trn}
\ , \label{RdotSEAFHDRO} \\
  \dot O &= (1-\fAF)\,{I\over \tao} + (1-\fFH){F\over \tfo}
\ . \label{OdotSEAFHDRO}
\end{align}
\end{subequations}

As in the SIR model, the SEIHCDRO, SEAFHCDRO, and SEAFHDRO models all assume
exponential probability 
distribution functions (pdfs) for the durations of the phases (besides S).
But while the SIR model involves a single exponential pdf for the duration of the
Infectious phase,  in our more complex models, the infectious phases are split into
several, each with an exponential pdf of duration, which effectively
creates a different pdf for the duration of the entire infectious phase,
i.e. the combination of (Infectious or Asymptomatic and Feverish),
Hospitalized (and Critical)
phases. Thus the pdf of the entire set of infectious
phases is no longer exponential, but the convolution of different
exponentials, which amounts to  gamma or Erlang distributions,
which both reach their maximum at a positive value of the duration.

\subsection{Assumptions and caveats}

The assumptions of the models presented above can be summarized as follows.
\begin{enumerate}
  \item The probability distribution functions of the durations for the
    transition from one phase to the next are all assumed to be exponential,
    but with mean durations that depend on the particular transition.
  \item Only fractions of the population advance to the next phase, hence the
    \emph{branching fractions} $f_{\rm A\to B}$, which all lie between zero and
    unity, and are assumed different.
  %% \item The transitions to recovery (the O phase), without the need for
  %%   hospitalization, have durations that also follow exponential pdfs, with
  %%   mean durations $\tio$ and  $\tfo$ for recovery from the Infectious or
  %%   Feverish stages, respectively.
  %% \item Similarly, the transitions to recovery from the hospital (which are
  %%   tabulated), also have durations whose pdfs are exponential, with mean
    %%   durations $\thr$ and $\tcr$.
    \item The free parameters are the \rz\ factors (before and during
      lockdown), the normalization of the first phase of Infectious
      (i.e. Asymptomatic in the latter two models), and 7 (SEIFHCDRO and
      SEAFHDRO) or 9 (SEAHCDRO) ratios of
      timescales to branching fractions.
    \item These free parameters are assumed to be fixed in time: i.e. the
      ratios of timescales to branching fractions are assumed fixed in time,
      and the \rz\ factor is assumed to vary as a step function at the start
      of the lockdown.
    \item The Feverish, Hospitalized, and Critical people are assumed to
      infect few others. The Feverish stay at home. The Hospitalized and
      Critical are assumed to infect a negligible number of Hospital staff.
      While this was not the case in the early phases of the pandemic in each
      country, where
      hospital staff were often infected by patients, my model assumes that
      by 1 March 2020 when the model starts, and especially starting on 18
      March 2020, where the hospital data begins, the  staff (doctors,
      nurses, and attendants) in French hospitals were sufficiently protected
      with face masks and with frequent hand washing and disinfection of
      their hands.
    \item People who recover, whether out or in the hospital, are assumed
      to immediately lose their infectiousness.
      A possible continual of infectiousness is possible, at least in some of
      the recovered people. The guidelines from the American Centers for
      Disease Control and Prevention (CDC) recommend 3 days of isolation after recovery
      \citep{CDC20}.
%%       This goes against
%% several reports, which claim instead that
%%         recovered patients remain Infectious for days to weeks.
         The modeling of a possible continuation of contagiousness in
         recovered people is straightforward (and has been
        performed by this author), but is beyond the scope of the present work.
\end{enumerate}

\section{Data}
\label{sec:data}

I used the daily COVID-19 hospital movements related to COVID-19 provided
by \emph{Sant\'e publique France}
at
https://www.data.gouv.fr/fr/datasets/r/6fadff46-9efd-4c53-942a-54aca783c30c.
This data lists the hospital arrivals $\dot {\cal H}$,
the arrivals in critical care (reanimation) $\dot {\cal C}$,
the deaths $\dot{\cal  D}$, and the releases
$\dot {\cal R}$.
These values are given for
the 96 French \deps, plus 5 overseas ones.
I only considered the 96 \deps, assuming that the population was
genetically homogeneous, leading to unique values of the mean durations of the
phases and of the branching ratios.
I either considered
the 96 \deps\ together as a single entity, called
`single-zone France', as well as
%% After testing my 3 models 
%% %I first tested my 3 models
%% using  rapid-to-run, few-zone models: France as a
%% single model, and models of 2, 4 and 8 \deps,
%% I ran models for 
15 regions, corresponding to the 13
French \emph{r\'egions}, but splitting \emph{r\'egion} \^Ile-de-France around the capital
into three units: Paris, \emph{Paris-Petite-Ceinture} (surrounding 3 districts:
Hauts de Seine [92], Seine-Saint-Denis [93], and Val de Marne [94]), and
\emph{Paris-Grande-Ceinture}
(outer districts:
Seine-et-Marne [77], Yvelines [78], Essonne [91], val d'Oise [95]).

%% or I considered 4 or 8 representative \deps, spanning
%% different levels of population.
The hospital data begins on 19 March 2020, and I used the values up to 3 May
2020, for a total of 46 days of data.\footnote{Some of the figures in
  Sect.~\ref{sec:GoF} display 
  more recent data.} 

I assumed that by number of hospital arrivals $\dot{\cal H}$, the hospitals
are including the number of arrivals into critical care.
%% In other words, we
%% expect
%% \begin{equation}
%%   \dot{\rm  H} = \dot {\cal H} - \dot {\rm C} \ .
%% \label{hospmodel1}
%% \end{equation}
But there were a few (1\%) cases where the hospital arrivals on a given date
is smaller than the arrivals in critical care.
In this case, I simply assumed
that the new hospitalizations do not include the entries
into critical care. 
%% In model $\rm H_{-1}$, I assumed zero new hospitalizations if the number of
%% arrivals in critical care exceeded that of hospitalizations.
%% Finally, I also included a model $\rm H_0$ that assumes that hospital
%% arrivals in the government data excludes the critical care arrivals.

\section{Analysis}
\label{sec:analysis}

%% Given the expected degeneracies between our $19+3\,N$ free parameters (12
%% mean durations, 7 branching ratios, and for each \dep\ the pre- and
%% post-lockdown $R_0$ factors as well as a normalization), rather than
%% find the best fit of these parameters,
%% I performed a Markov Chain Monte
%% Carlo (MCMC) exploration of the parameter marginal distributions and
%% covariances.
%% Working on a single region, with 37 days of data, i.e. 148 data points, I
%% thus have 21
%% free parameters.
%% One can combine $N$ different regions with similar genetic makeup, in which
%% case one can assume that the 12 mean durations and 7 branching fractions are
%% the same for all $N$ regions, leaving only 3 free parameters per region: the
%% two values of $R_0$ and the initial fraction of Asymptomatic people. This would
%% lead to $19+ 3\,N$ free parameters for $4\,N\,d$ data points, where $d$ is the
%% number of days of data.

\subsection{Initial conditions}
At the origin of time, 1 March 2020, I set all phases $X=0$, except the first
Infectious phase ($I$ or $A$) which is given a small value (whose logarithm is a free
parameter) and the Susceptibles which initially satisfy $S = 1-I$ or $S=1-A$,
depending on the model. The date of 1 March 2020 was not chosen at random,
but corresponds to 3 days after the start of the exponential growth in
France. Some regions only saw exponential growth several days later, so 1
March was chosen as a compromise between the early regions (around Mulhouse
in the Grand-Est) and the late regions (west of France).

\subsection{Free parameters}

As mentioned above, the free parameters are two \rz\ factors (before and
after lockdown), the normalization of the earliest Infectious phase, and 7 or
9 ratios of timescales to branching fractions.
I have analyzed France as a single zone in this manner, and the computations
are rapid. But one can obtain a tighter constraint on the ratios of
timescales to branching fractions by joint modeling several zones of France,
all assumed to be genetically homogeneous, i.e. with the same
time-over-fraction ratios, but allowing for different \rz\ factors before and
during lockdown, as well as different normalizations.
The total number of free parameters were then
$N_{\rm time-over-fraction} + 3\,N_{\rm zones}$, i.e. up to 54 free
parameters (for 15 regions in the SEAFHCDRO model).
%% I have run models with single-zone (France or Paris), 2, 4, and 8 \deps, as
%% well as 15 regions: the 13 national regions, but with the Paris region (\^Ile
%% de France) split into Paris itself, the inner suburbs (\deps\ 92, 93, and 94)
%% and the outer suburbs (\deps\ 78, 91, 95).
%% Finally, I managed to run very time-costly cases with all 96 \deps\ of
%% metropolitan France (excluding the overseas \deps \ and territories, to
%% achieve better genetic homogeneity).
%% \textcolor{red}{But I currently forced $R_0^{\rm lockdown}$ to be the same
%%   for all 96 \deps.}

\subsection{Likelihood}

Given the hospital data
\begin{equation}
  \dot{\cal X}_i \equiv \left\{
  \dot {\cal H},
       [\dot{\cal C},]
       \dot {\cal D},
       \dot {\cal R} \right\} \
       ,
\end{equation}
the likelihood of my models is written
\begin{equation}
-\ln {\cal L} = -\sum_{i=1}^{N_{\rm zones}}\,
 \sum_{j=1}^{N_{\rm times}}\,
 \sum_{k=1}^{N_{\rm phases}}
\ln {\rm Poisson}\left(\dot {\cal X}_k (t_j,i) \,|\, 
    {\rm Population}_i\,f_{i,k}\left(t_j,\right) \right)\ ,
\end{equation}
where
$f_{i,k}\left(t_j\right)$ represents the predicted number of daily arrivals into
phase ${\cal X}_k$ in zone $i$ on day $j$.

\subsection{MCMC analysis}
\begin{table}[ht]
  \begin{center}
  \caption{Priors on free parameters}
    \begin{tabular}{lrr}
      \hline
      \hline
      Parameter & minimum & maximum \\
      \hline
      fractions & 0$\quad\ $ & 1$\quad\ $\\
      log (durations/days) & 0$\quad\ $ & 2$\quad\ $ \\
      $R_0$ before lockdown ($R_0^{\rm Ini})$ & 1$\quad\ $ & 6$\quad\ $ \\
      $R_0$ after lockdown ($R_0^{\rm Conf})$ in units of $R_0^{\rm Ini}$ & 0$\quad\ $ & 1$\quad\ $ \\
      log fraction of Asymptomatics on 1 March 2020 & --8$\quad\ $ &
      --3$\quad\ $ \\
      \hline
     \end{tabular}
    \parbox{0.73\hsize}{{\sc Notes}: The logarithms are in base 10.
      % \textcolor{red}{I changed to relative $R_0^{\rm confine}$.}
}
  \end{center}
  \label{tab:priors}
\end{table}

Marginal distributions of the free parameters and of the evolution of the
pandemic in France were obtained through a Markov-Chain Monte Carlo (MCMC) model.
The code was written in Python 3, and the marginal distributions of the free
parameters were estimated
from the posteriors using the {\sc emcee} package \citet{Goodman&Weare10}.
The MCMC procedure involved a minimum of number of chains (or
``walkers'') equal to twice the number of free parameters, with a minimum of
64 (for single-zone models), and up to 119 for 15-region models, and up to
near 600 for 96-department runs.
The chains (walkers) are advanced using the \emph{stretch move}
algorithm  \citep{Goodman&Weare10}.
These chains were initialized with uniform sampling over the allowed range of
parameters shown in Table~\ref{tab:priors}.
The chains were run for 200\,000 steps (single zone) or 50\,000 steps (15
regions),
thus involving a total of several 
million evaluations of the set of differential equations.
%% ~(\ref{Sdot})   to
%% (\ref{Odot2}).
%% The fraction of Neutralized is deduced with equation~(\ref{sum}).
I assumed 80\% \emph{burnin} factor, thus only considering
the last 20\% of each chain
%(i.e. last 10\,000 steps)
for the analysis.
The figures showing curves, use medians for the curves with shaded regions
for 16-84\% and 5-95\% confidence levels.

\section{Results}
\label{sec:results}

%% Paris (75),
%% Haut-Rhin (68, that with the highest fraction of deaths, encompassing
%% Mulhouse),
%% as well as
%% Bouches-du-Rh\^one (13, including Marseille),
%% Creuse (23, a low-density \dep),
%% Eure (27, low density),
%% Gironde (33, including Bordeaux),
%% Sarthe (72, including Le Mans),
%% and
%% Seine-Maritime (76, including Rouen and Le Havre).

%\subsection{Goodness-of-fits of single-France and 15-region models}
\subsection{Goodness-of-fits}
\label{sec:GoF}

\begin{figure}[ht]
  \centering
  \includegraphics[width=0.4\hsize]{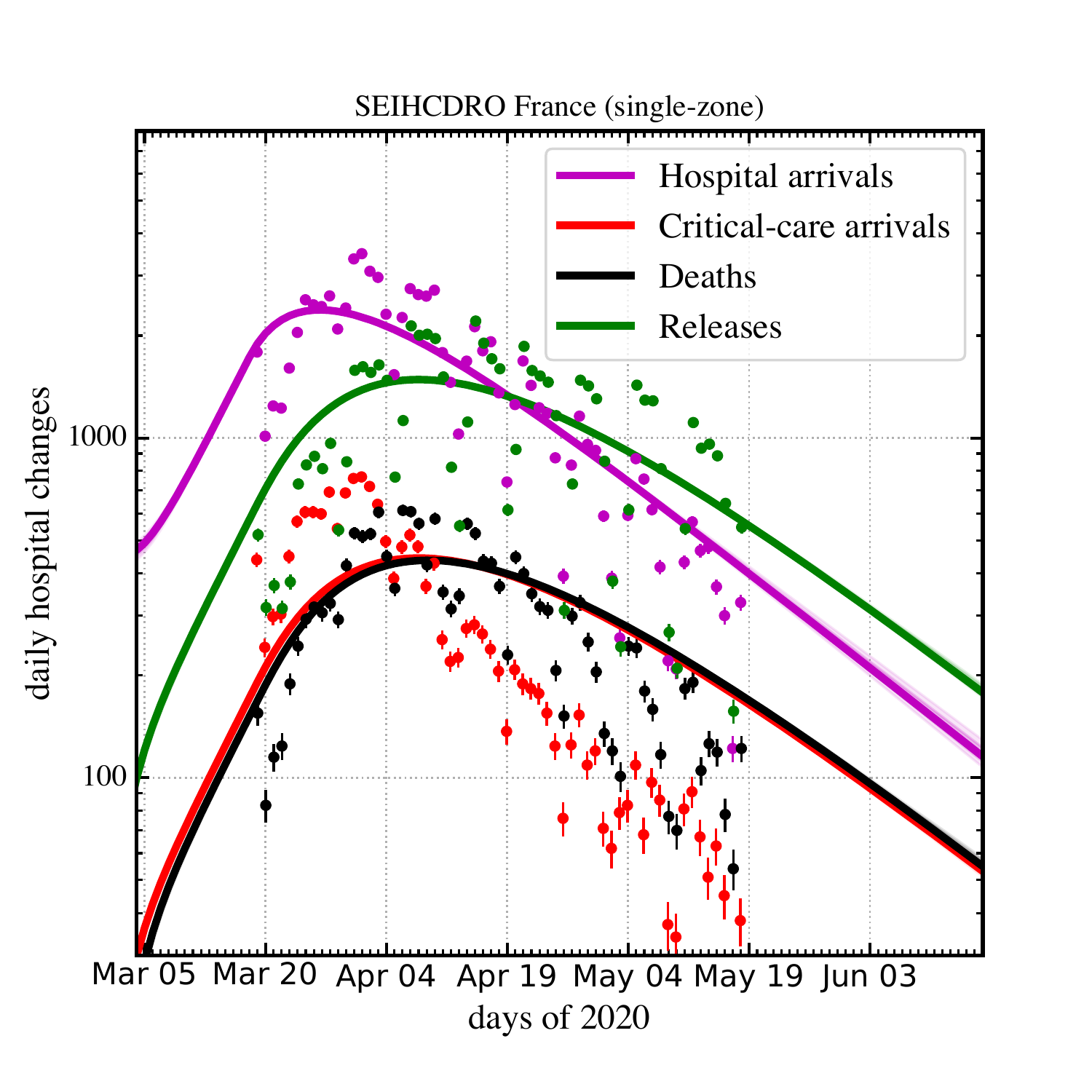}
  \includegraphics[width=0.4\hsize]{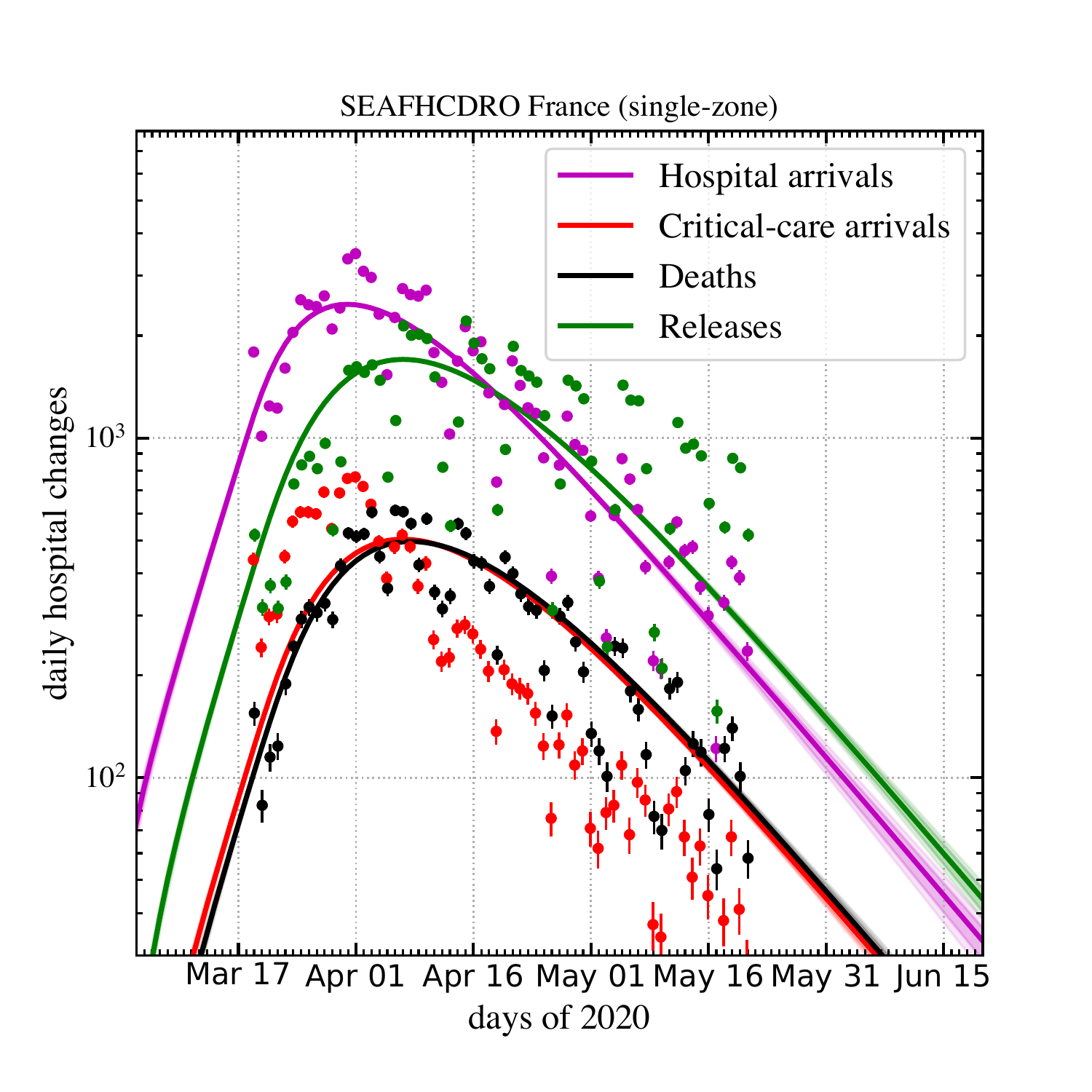}
  \caption{
Goodness-of-fit results for single-zone France. The \emph{symbols} show the
daily hospital data (arrivals in \emph{magenta},
arrivals in critical care in \emph{red},
deaths in \emph{black},
released in \emph{green}),
with Poisson error bars. The \emph{curves} are the median predictions from the
SEIHCDRO (\emph{left}) and SEAFHCDRO (\emph{right}) models (both fitting to
data up to 3 May), while the
narrow and wider \emph{shaded areas} respectively
show the extent of 16th-84th and
5th-95th percent confidence levels.
  }
\label{fig:GoFFrance}
\end{figure}

Figure~\ref{fig:GoFFrance} shows the goodness of fit of single-zone France for the SEIHCDRO and
SEAFHCDRO models. The daily hospitalizations (magenta), deaths (black) and
releases (green) are well fit by the models, especially for the SEAFHCDRO
model (right), while the SEIHCDRO model predicts too early a peak in daily
hospital arrivals. However, the daily arrivals in
critical care are poorly fit by the models, which cannot reproduce the early
narrow peak in the last days of March, followed by the rapid decrease.
One also notices that the mean values of the post-burnin marginal
distributions of \rz\ lead to very high values before the lockdown:
$R_0 = 6$ (the allowed upper limit)
for model SEIRHCDRO and $R_0=4.24\pm0.1$ for model SEAFHCDRO.
At the same time, the values during lockdown are
found to be very small: $R_0 = 0.18\pm0.02$ and $0.01\pm0.02$ for SEIRHCDRO and
SEAFHCDRO, respectively.

\begin{figure}[ht]
  \centering
  \includegraphics[width=0.4\hsize]{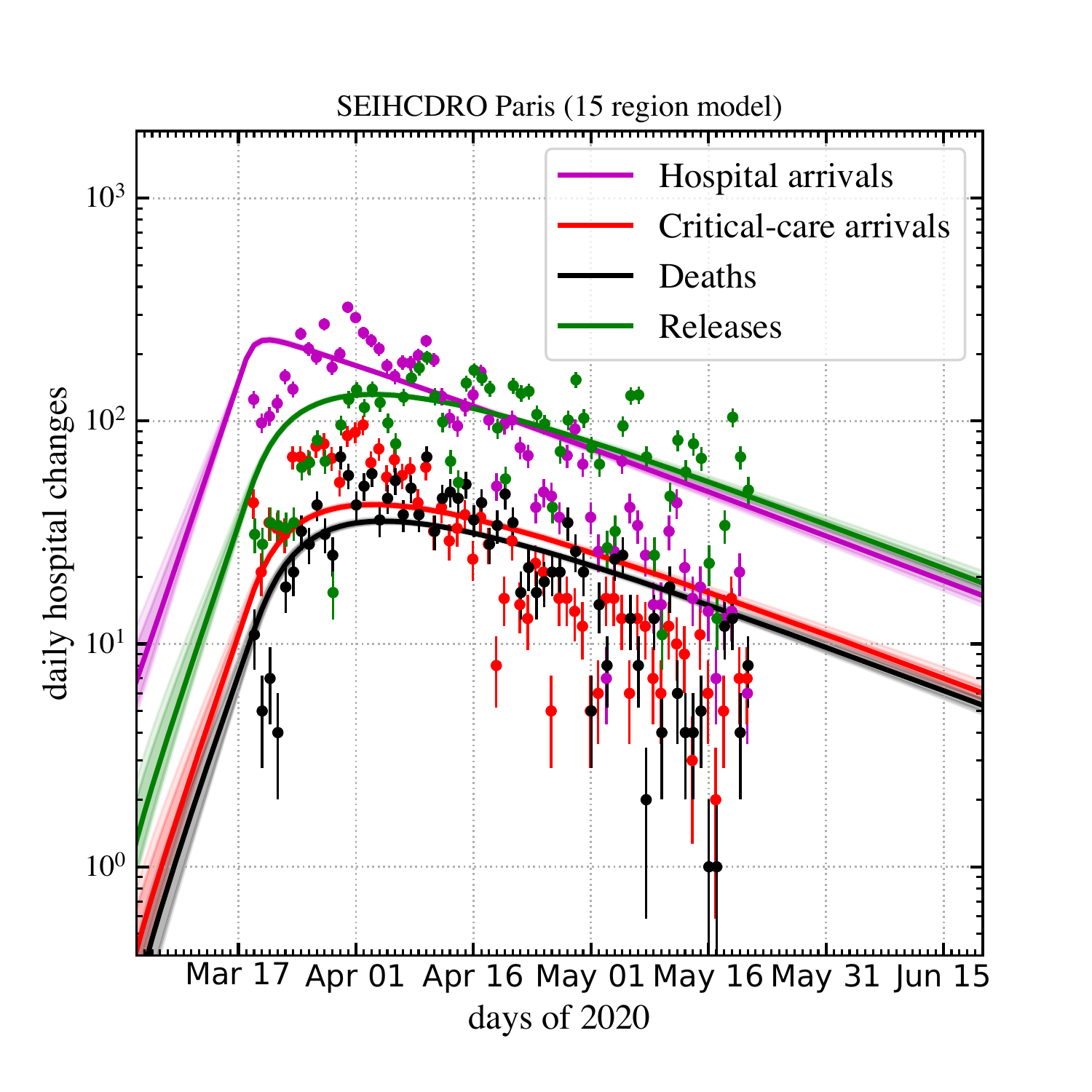}
  \includegraphics[width=0.4\hsize]{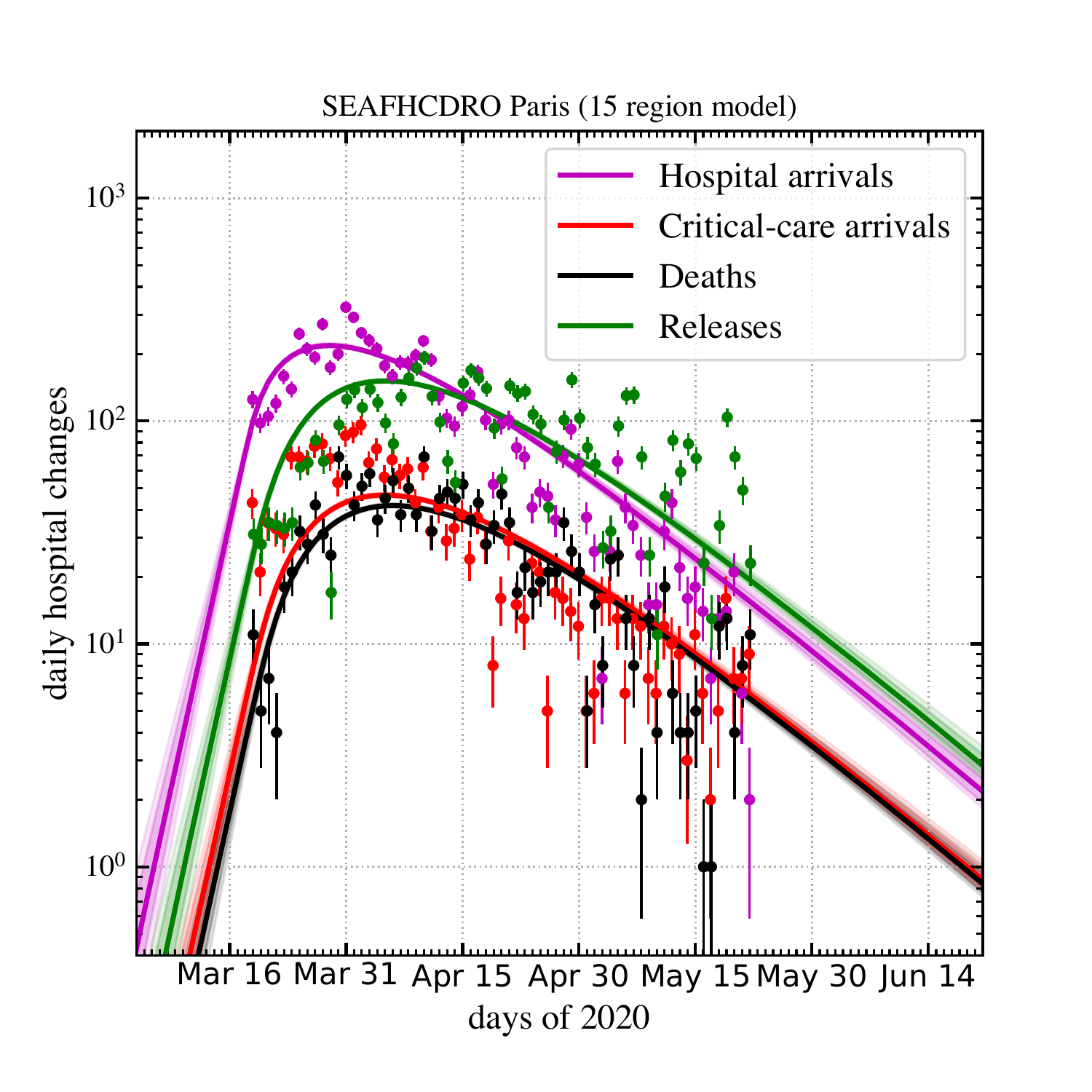}
  \caption{
    Same as Figure~\ref{fig:GoFFrance}, again for the SEIHCDRO (\emph{left}) and
    SEAFHCDRO (\emph{right}) models,
    but for Paris from the 15-region analysis.
  }
\label{fig:GoFRegionsParis}
\end{figure}

Figure~\ref{fig:GoFRegionsParis} shows the same trends for the city of Paris,
when running models  SEIHCDRO (left) and
SEAFHCDRO (right) with the 15 regions.
One notices the same poorly-fit data for the arrivals in critical care as for
all of France. Since France contributes roughly 10\% of France in terms of
critical care arrivals, this suggests that the dichotomy between model
predictions and data for critical care arrivals is general throughout France
(not shown in the figures).
This dichotomy with the models cannot be caused by saturated critical care,
because it occurs in all regions, while only a few suffered from critical
care unit saturation.
Interestingly, a hospital official mentioned that France has learned how to
better treat COVID-19 patients and has been sending progressively smaller
fractions to critical care.\footnote{Radio broadcast, \emph{Europe 1}, 5 May 2020,
  around 9AM.}

With joint modeling of 15 regions, the \rz\ values before and during lockdown
are respectively lower and higher 
than for single-Zone France, suggesting that the lockdown is less effective than
for single-zone France, where  \rz\ is reduced by a factor of at least 30.
For Paris, the \rz\ factors before lockdown
are
$R_0 = 2.65\pm0.20$ for SEIHCDRO and
$R_0 = 3.5\pm0.2$ for SEAFHCDRO.
At the same time, the lockdown values of \rz\ is not as low for Paris than
for the single-zone France model:
$R_0 = 0.91\pm0.05$ and $R_0 = 0.65\pm0.20$ for models SEIHCDRO and
SEAFHCDRO, respectively.

\begin{figure}[hbt]
  \centering
  \includegraphics[width=0.4\hsize]{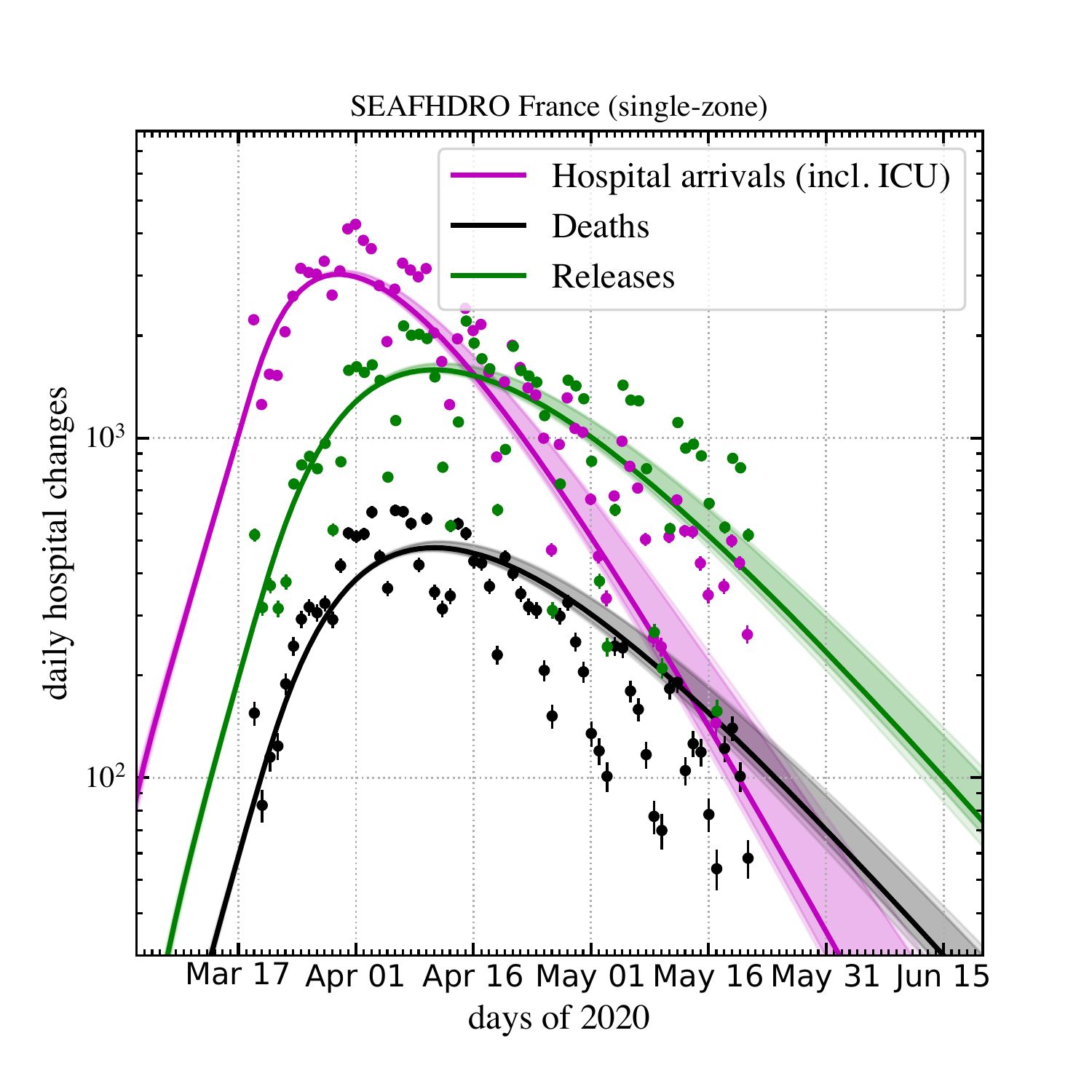}
  \includegraphics[width=0.4\hsize]{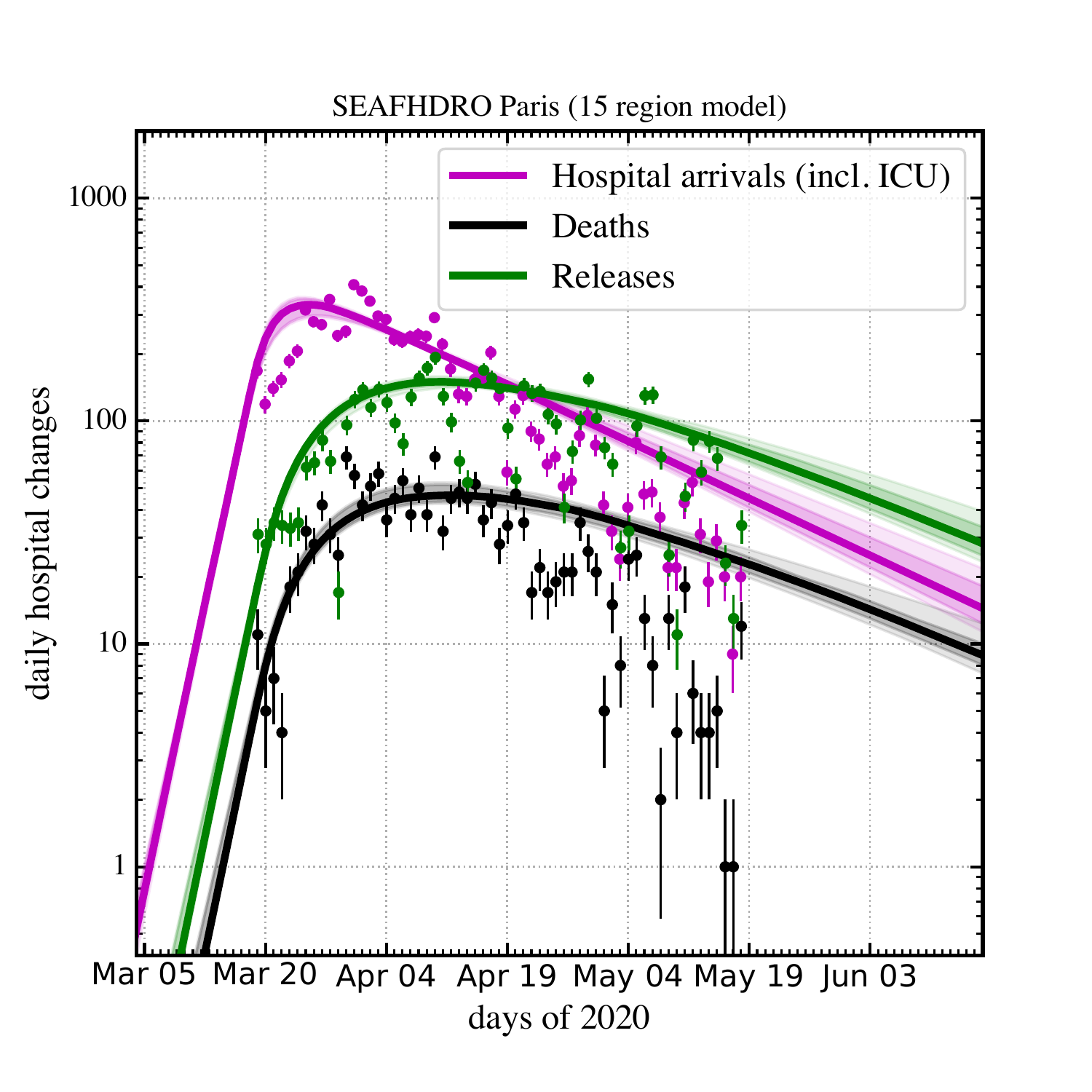} 
  \caption{
Goodness-of-fit for the SEAFHDRO model (i.e. with Critical phase merged into
Hospitalized) for single-zone France (\emph{left}) and for Paris in the
15-region run (\emph{right}), both fitting data up to 10 May.
  }
\label{fig:GoFm8}
\end{figure}

The comparison of the quality of the fits between the SEIHCDRO and SEAFHCDRO
models is performed using BIC Bayesian evidence
\begin{equation}
  {\rm BIC} = -2\,\ln {\cal L}_{\rm MLE} + N_{\rm
    parameters} \, \ln N_{\rm data}\ ,
\end{equation}
where ${\cal L}_{\rm MLE}$ is the maximum likelihood.
BIC effectively compares the log-likelihoods, but penalizes extra free
parameters, especially in large data sets (as is the case here).
Such a test can only be done on the same amount of data, which is the case
for these two models.
In all cases studied, including single-zone France, the SEAFHCDRO has  lower BIC
than SEIRHCDRO, by typically over 1000, whereas a difference of 6 is deemed
strong evidence.

The better fits of the SEAFHCDRO model led me to drop the SEIHCDRO model.
Moreover, the general dichotomy between the predicted and observed arrivals to critical
care led me to also consider the SEAFHDRO model, which folds the Critical
phase into the Hospitalized phase.

Figure~\ref{fig:GoFm8} shows the goodness-of-fits of the SEAFHDRO model for
the single-Zone France as well as for Paris in the 15-region run.
The fits remain good for the daily Hospital arrivals (magenta) and the
hospital Releases (green), but now fail somewhat to describe the decrease in
the daily deaths (black). Moreover, even if one cannot use the BIC Bayesian
evidence, because the number of data points is not the same, we note that a
na\"{\i}ve correction of the $-\cal {L}_{\rm MLE}$ by 4/3 for SEAFHDRO leads
to a value 18 above that of SEAFHCDRO, 

All models used here predict that the daily Deaths should evolve in parallel
to the daily Releases, with a possible time lag, whereas the data indicates
that the Deaths fall more rapidly than the Releases (see
Figs.~\ref{fig:GoFFrance}, \ref{fig:GoFRegionsParis}, and \ref{fig:GoFm8}).

\subsection{Timescales and branching fractions}

\begin{figure}[ht]
  \centering
  \includegraphics[width=0.49\hsize]{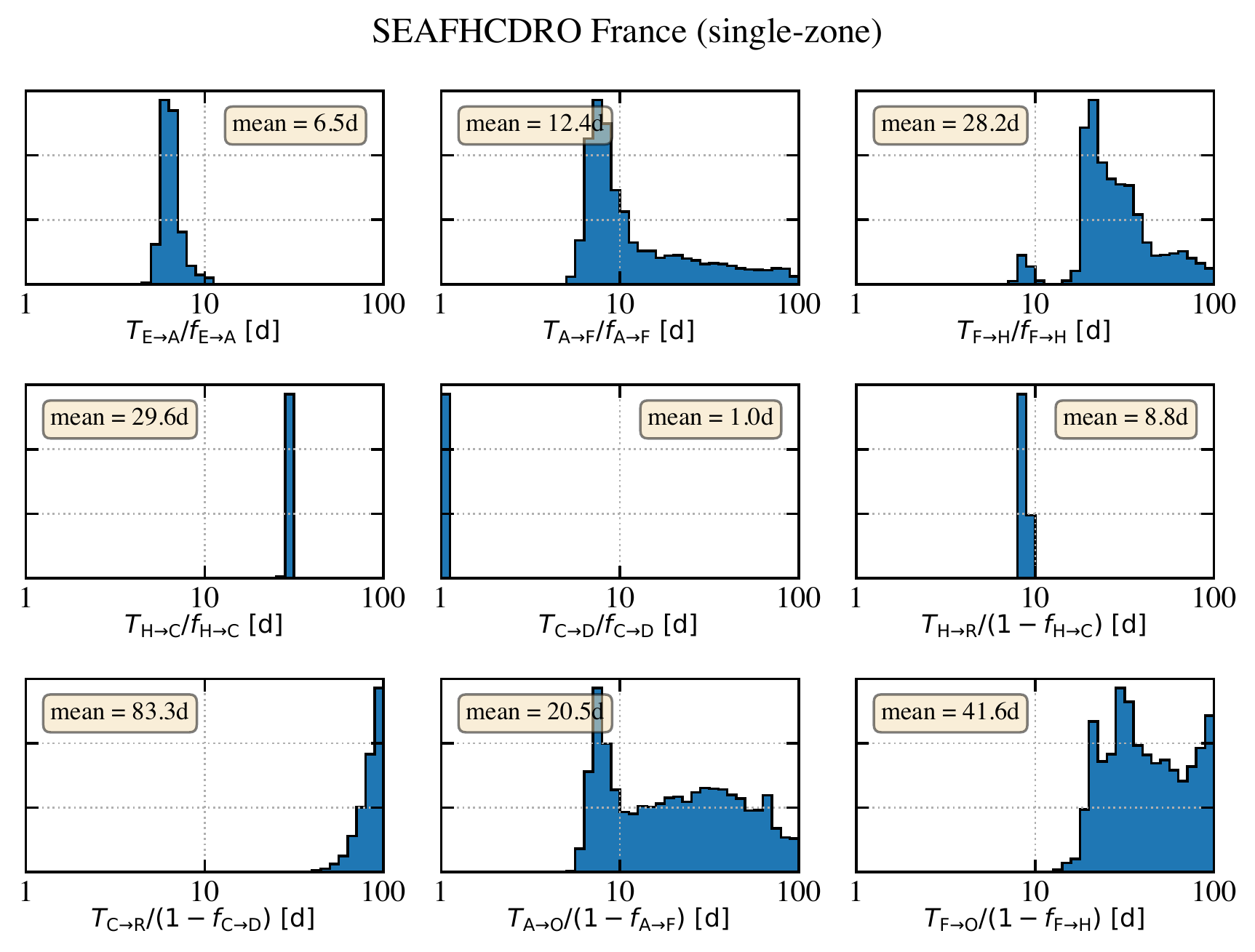}
  \includegraphics[width=0.49\hsize]{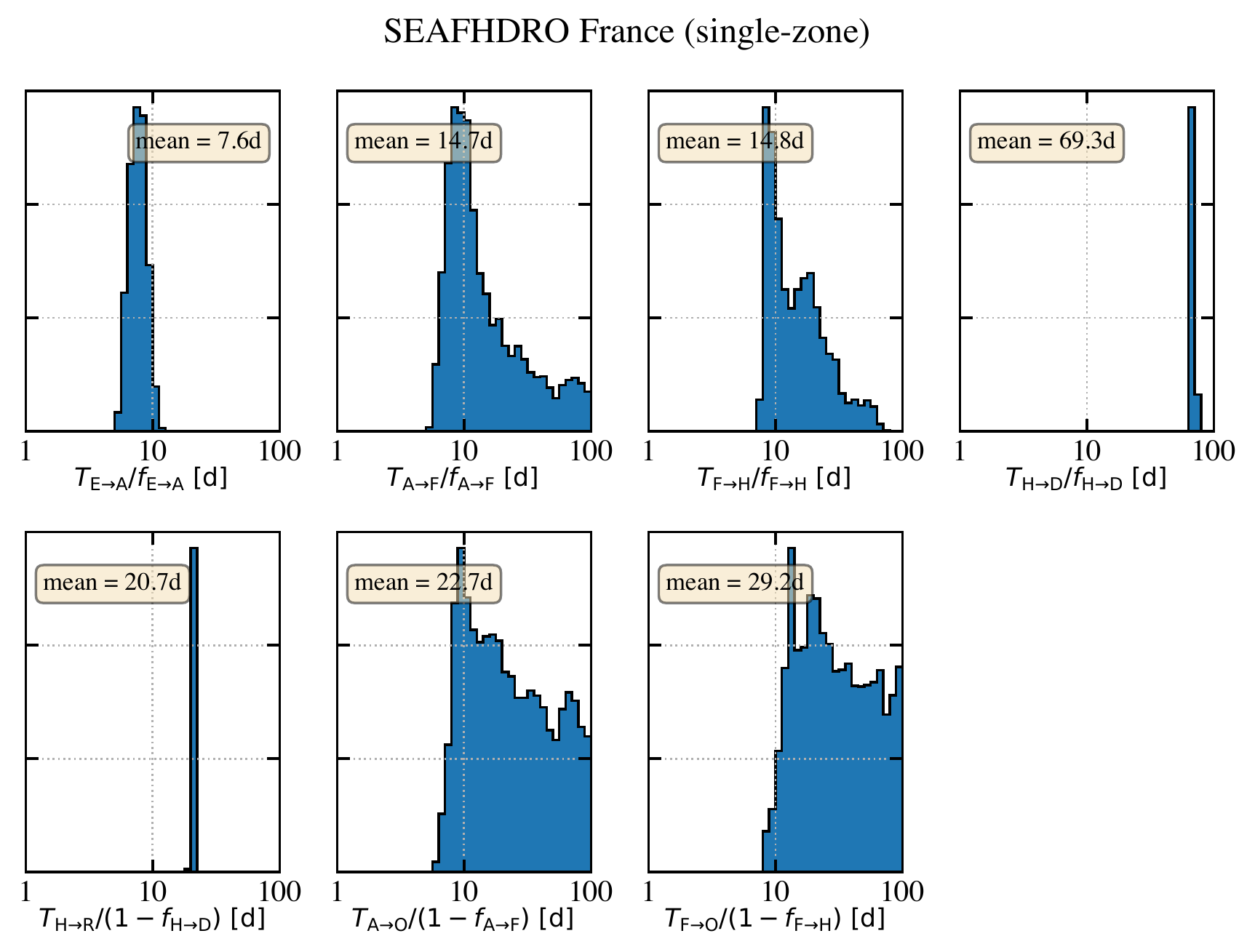}
  \includegraphics[width=0.49\hsize]{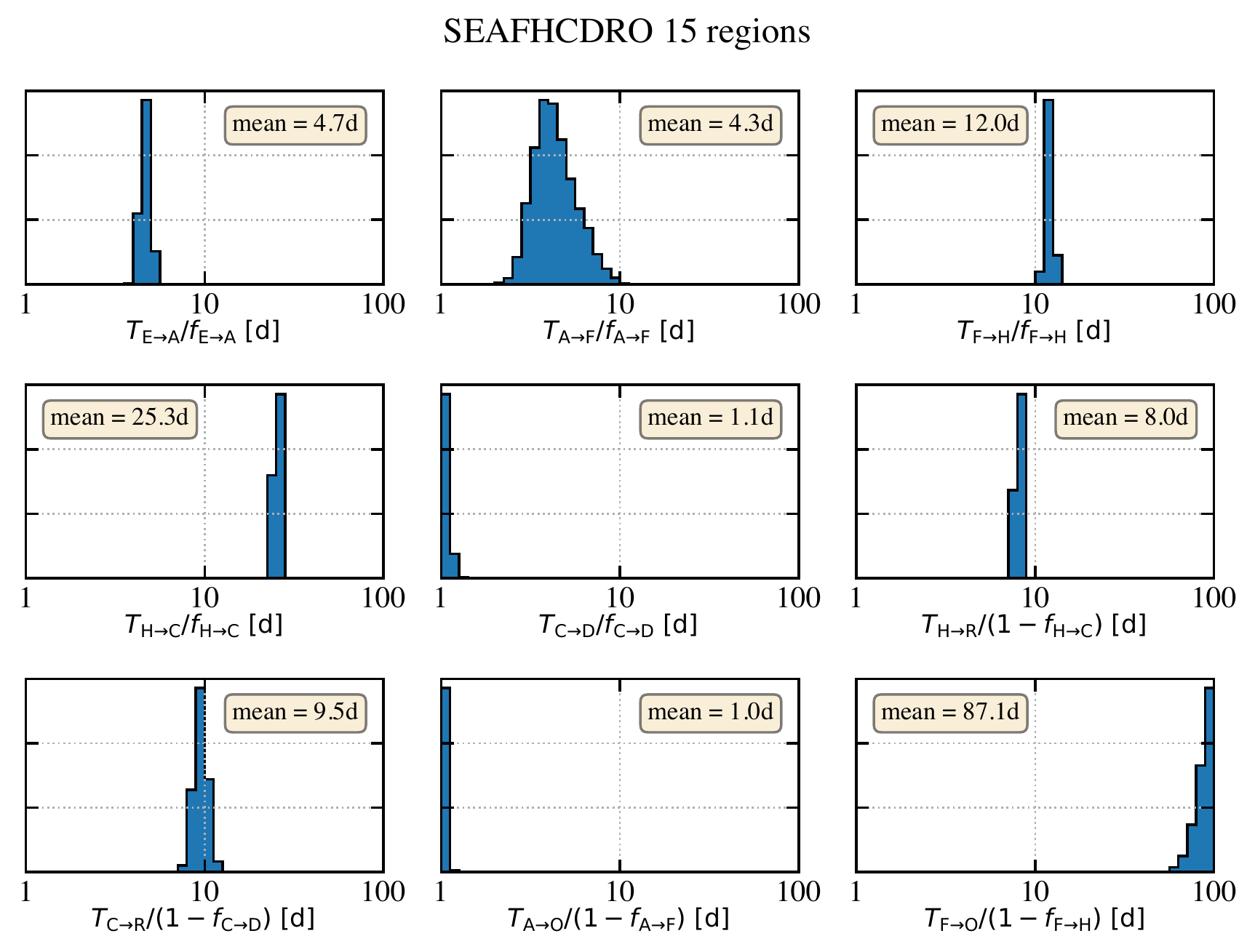}
  \includegraphics[width=0.49\hsize]{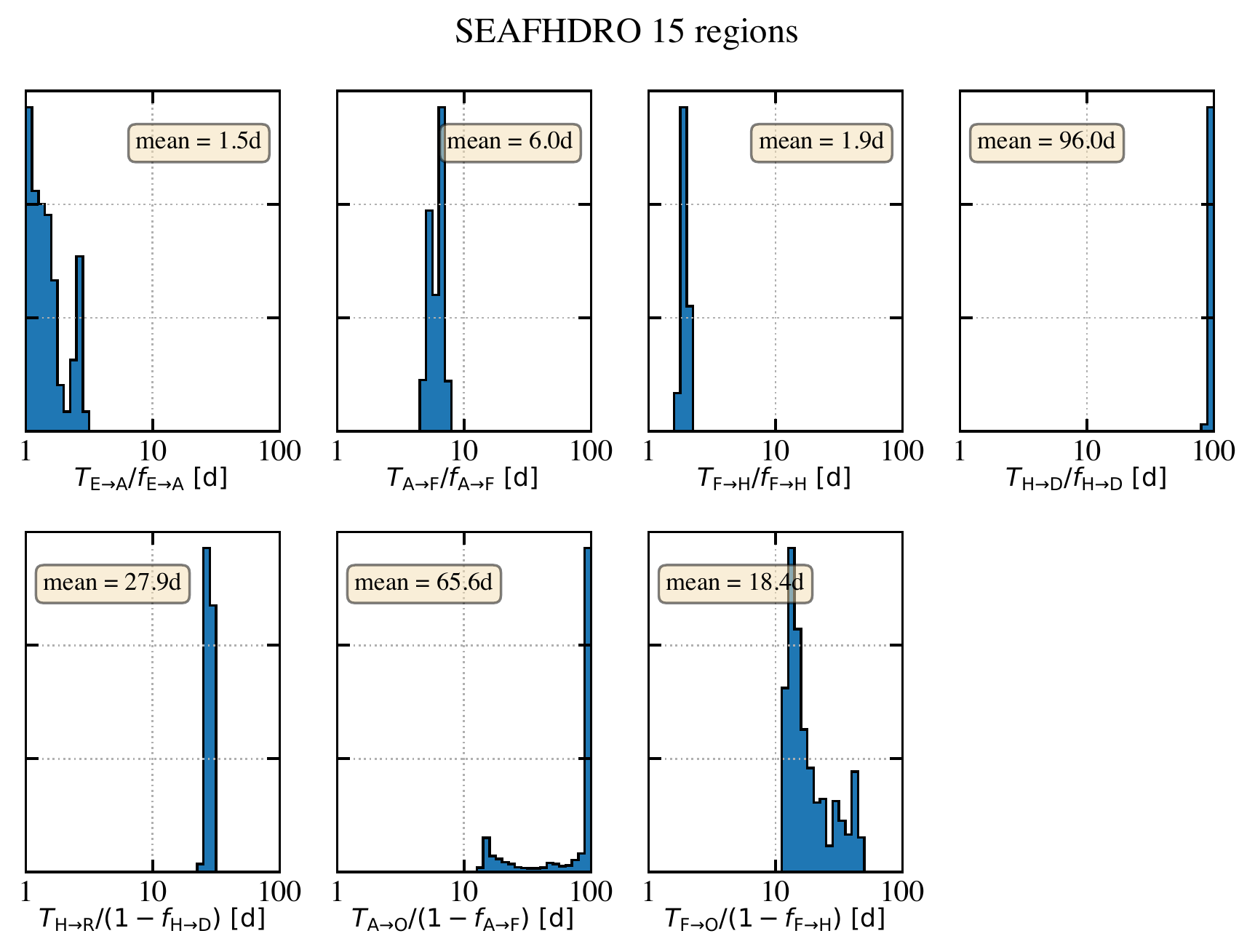} 
  \caption{
Marginal distributions of timescales over branching fractions for the
single-zone France ({\rm top}) and 15-region ({\rm bottom}) runs for models
SEAFHCDRO (\emph{left}) and SEAFHDRO (\emph{right}).
The \emph{boxes} indicate the geometric mean timescales over branching fractions.
  }
\label{fig:margRatios}
\end{figure}
Figure~\ref{fig:margRatios} displays the marginal distributions of ratios of
timescales to branching fractions for the SEAFHCDRO (left) and SEAFHDRO
(right) models (with two fewer ratios), for single-zone France (top) and 15 regions (bottom).
One first notes that the joint fit of 15 regions allows much narrower constraints on
the ratios than the single-zone France fit.
This is most striking for ratios
$\tfh/\fFH$ and
$\tao/(1-\fAF)$,
whose marginal distribution go from very wide (single-zone France) to very
narrow (15 regions).

Several results can be inferred from these marginal distributions of ratios of
timescales to branching fractions.
In particular, since the timescale is necessarily less than the ratio,
Figure~\ref{fig:margRatios} provides upper limits on timescales.
The top left panel of the 15-region models (lower sets of panels) indicates that the
\emph{incubation time}, $\tea$, is at most a few days (95\% confidence upper limits
of 5.1 and 2.7 days according to models SEAFHCDRO and SEAFHDRO, respectively).
The time for strong symptoms to appear, $\taf$, is shorter than 7.1 days
(95\% c.l.).
The time to recover from the Feverish phase, $\tfo$ is large, or else the
fraction of Feverish that end up Hospitalized is near unity, which seems less likely.

Several marked differences appear between the two models:
The ratio $\tfh/\fFH$ is less than 2 days for SEAFHDRO in contrast to less
than 13 days for SEAFHCDRO (both at 95\% c.l.).
And the time for Asymptomatics to recover,
$\tao$, is less than 1.1 days (95\% c.l.) for SEAFHCDRO, but possibly as long as 100 days
for SEAFHDRO.
Finally, while the time from Critical to Death is shorter than 1.2 days (95\%
c.l.) in the
SEAFHCDRO model, the ratio $\thd/\fHD$ is peaked at 100 days (the allowed
maximum, see Table~\ref{tab:priors}). While these two measures are not directly comparable, this may mean
that only a small fraction of hospitalized eventually die in the hospital.

\subsection{The effects of lockdown on basic reproduction factors}

\begin{figure}[ht]
  \centering
  \includegraphics[width=0.47\hsize]{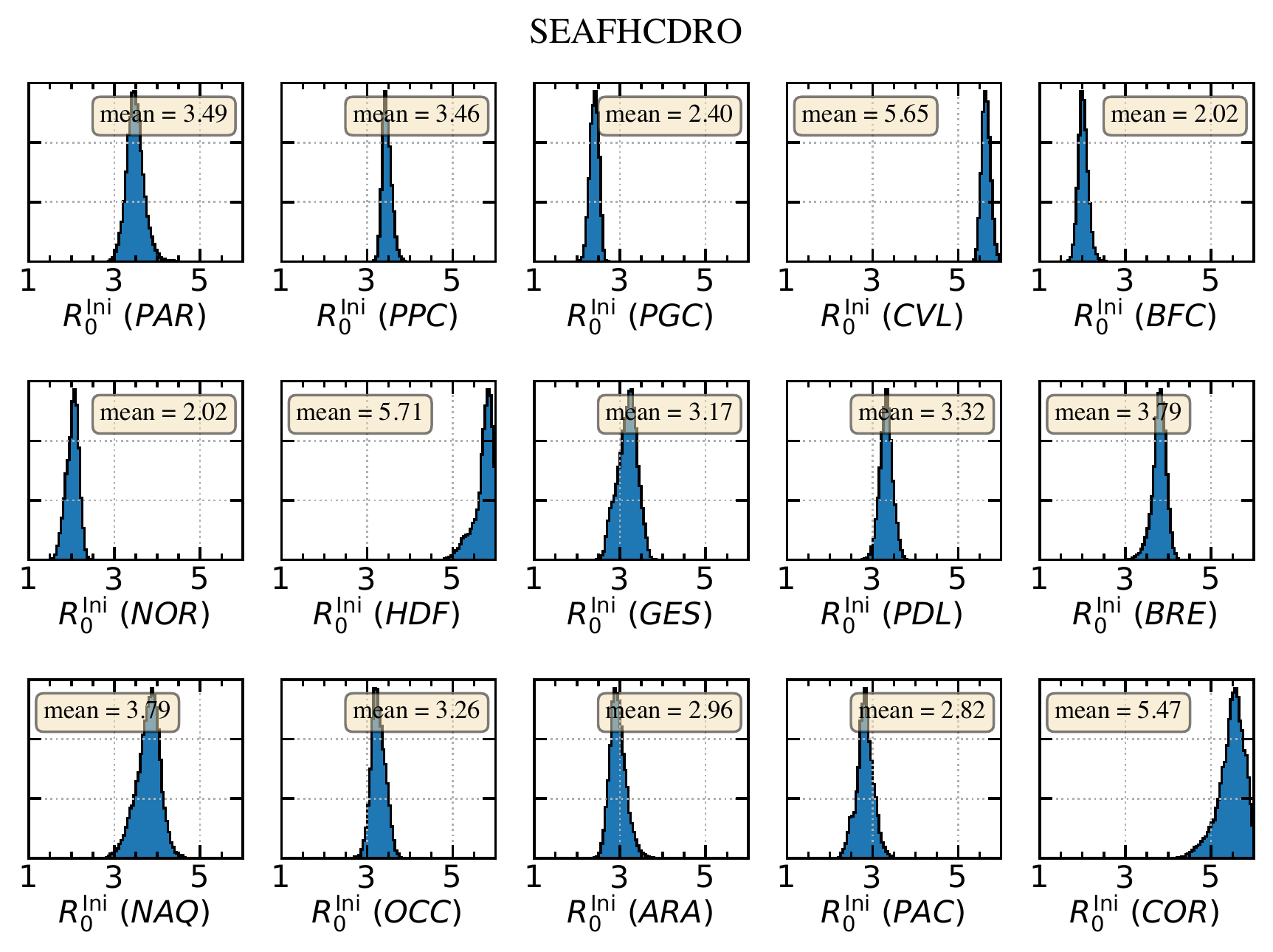}\qquad
  \includegraphics[width=0.47\hsize]{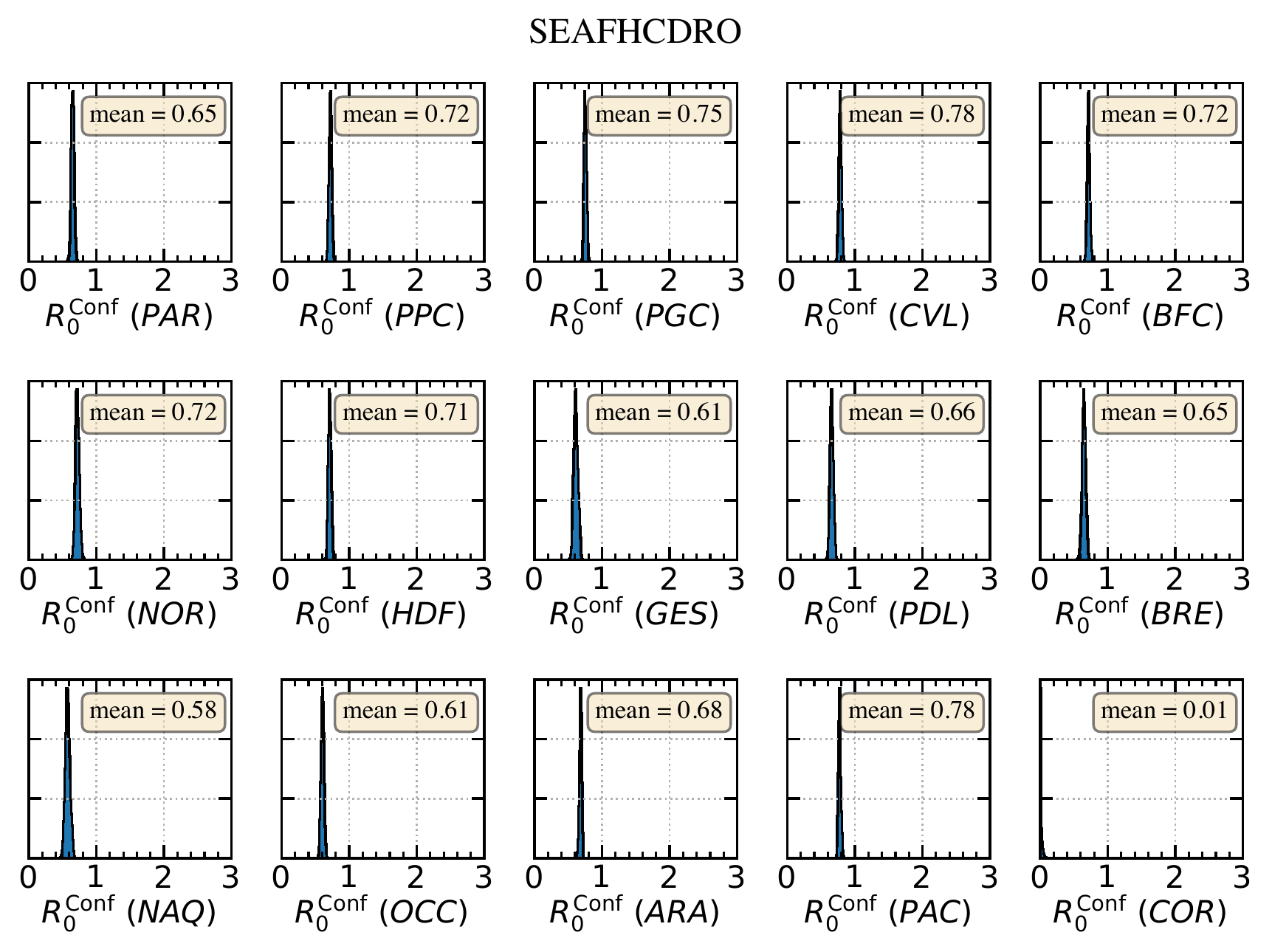}
  \includegraphics[width=0.47\hsize]{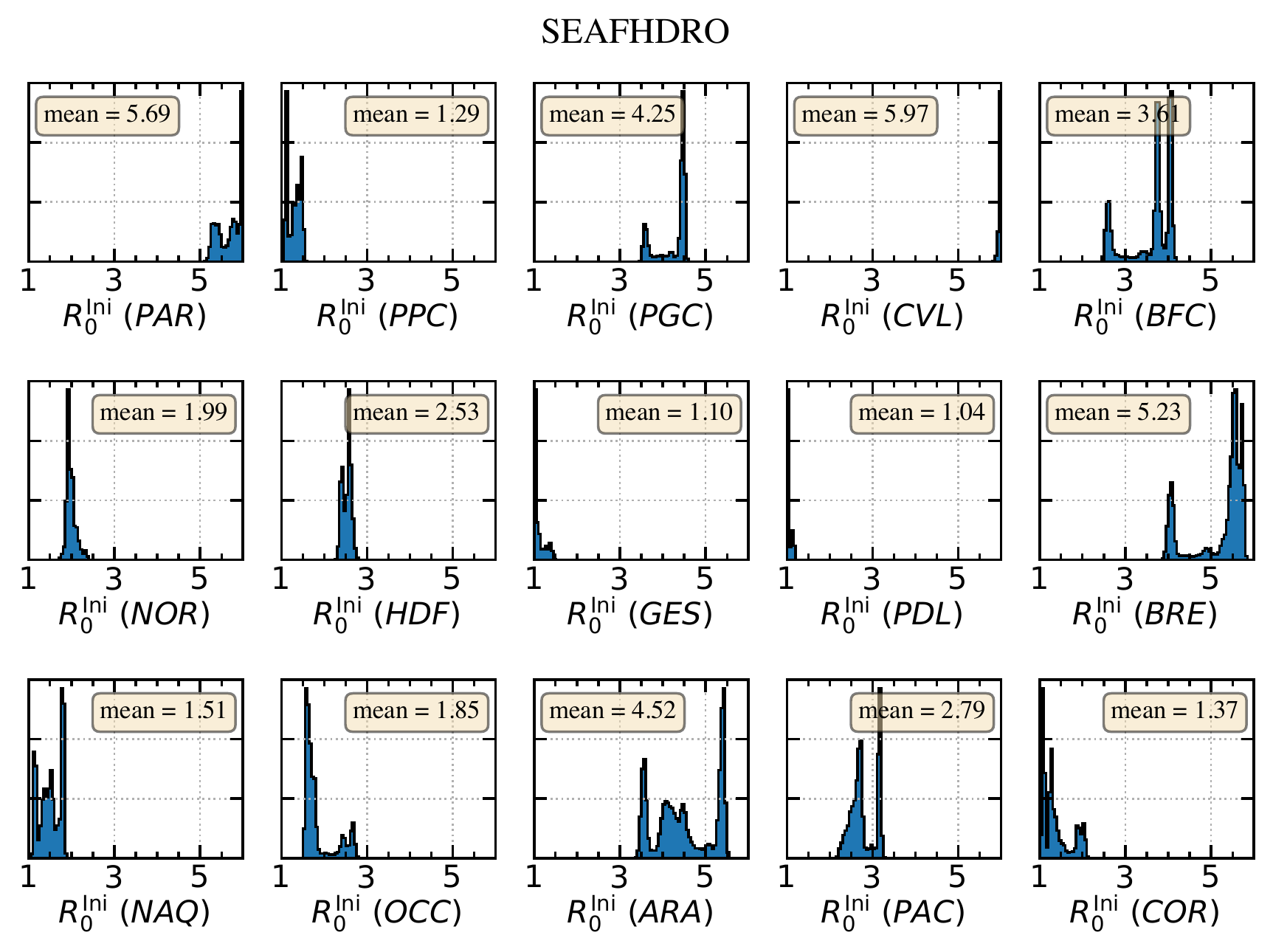}\qquad
  \includegraphics[width=0.47\hsize]{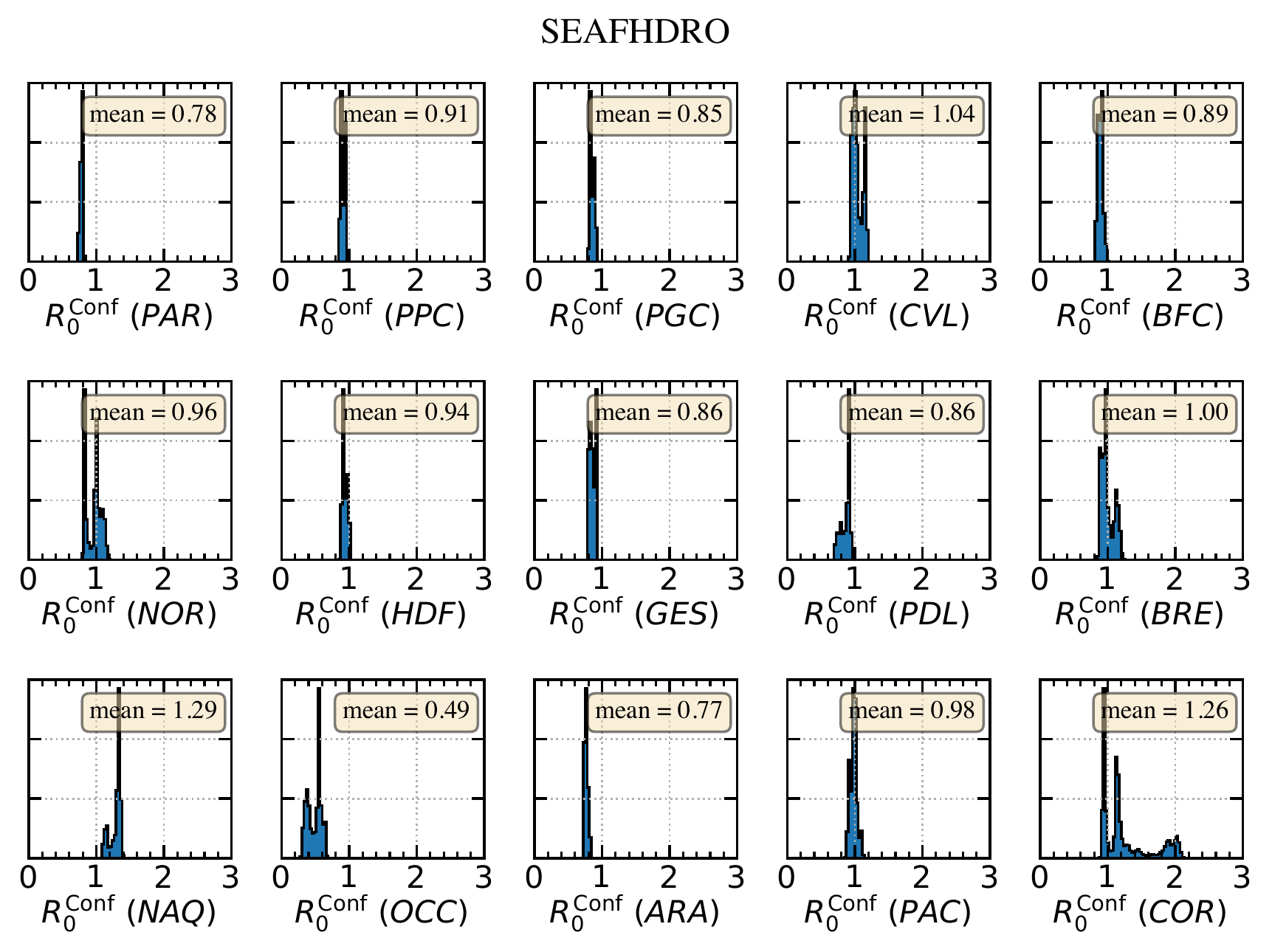}
  \caption{
    Marginal distributions of basic reproduction factors of 15 French regions
    before (\emph{left})
and during (\emph{right}, with shifted scale) lockdown for models SEAFHCDRO (\emph{top}) and
SEAFHDRO (\emph{bottom}).
The $x$-axis limits reflect the range of priors, except for the bottom right
panel, where the free parameter was $R_0^{\rm Conf}/R_0^{\rm Ini}$, whose
prior was flat between 0 and 1.
  }
\label{fig:margR0}
\end{figure}

Figure~\ref{fig:margR0} shows the marginal distributions of $R_0$ for the 15
regions of France before
(left) and during (right) lockdown for the SEAFHCDRO (top) and SEAFHDRO
(bottom) models.
Table~\ref{tab:R0} provides means and 90\% confidence limits for each region
and also provides mean values weighted by the populations of the regions.
Of course, the \rz\ factors during the lockdown were  well below the values
before the lockdown. This means that the lockdown had a noticeable effect on
the reproduction factor.

\begin{table}[ht]
  \caption{$R_0$ values by region}
  \begin{center}
    {\small
    \begin{tabular}{llccccc}
      \hline
      \hline
      Abb. & Name & \multicolumn{2}{c}{$R_0$ (before lockdown)} & &
      \multicolumn{2}{c}{$R_0$ (during lockdown)} \\ 
      \cline{3-4}
      \cline{6-7} 
      & & SEAFHCDRO & SEAFHDRO & & SEAFHCDRO & SEAFHDRO \\
      \hline
PAR & Paris &  3.49\ [3.17:3.85] & 5.69 \ [5.27:5.99] & & 0.65\ [0.62:0.69] &
 0.78\ [0.74:0.81] \\
PPC & Paris-Petite-Ceinture&  3.46\ [3.31:3.65] & 1.29 \ [1.10:1.50] & & 0.72\ [0.69:0.75] &
 0.91\ [0.86:0.96] \\
PGC & Paris-Grande-Ceinture &  2.40\ [2.23:2.56] & 4.25 \ [3.57:4.53] & & 0.75\ [0.72:0.78] &
 0.85\ [0.81:0.90] \\
CVL & Centre - Val de Loire &  5.65\ [5.49:5.82] & 5.97 \ [5.91:6.00] & & 0.78\ [0.75:0.82] &
 1.04\ [0.95:1.17] \\
BFC & Bourgogne-Franche-Comt\'e &  2.02\ [1.84:2.21] & 3.61 \ [2.57:4.09] & & 0.72\ [0.69:0.75] &
 0.89\ [0.84:0.95] \\
NOR & Normandie &  2.02\ [1.77:2.24] & 1.99 \ [1.86:2.21] & & 0.72\ [0.67:0.77] &
 0.96\ [0.82:1.12] \\
HDF & Hauts de France&  5.71\ [5.23:5.96] & 2.53 \ [2.37:2.67] & & 0.71\ [0.68:0.75] &
 0.94\ [0.89:0.99] \\
GES & Grand Est &  3.17\ [2.77:3.53] & 1.10 \ [1.00:1.39] & & 0.61\ [0.56:0.67] &
 0.86\ [0.80:0.91] \\
PDL & Pays de Loire &  3.32\ [3.07:3.56] & 1.04 \ [1.00:1.16] & & 0.66\ [0.62:0.70] &
 0.86\ [0.72:0.93] \\
BRE & Bretagne &  3.79\ [3.49:4.03] & 5.23 \ [4.02:5.75] & & 0.65\ [0.60:0.69] &
 1.00\ [0.89:1.17] \\
NAQ & Nouvelle Aquitaine &  3.79\ [3.28:4.20] & 1.51 \ [1.12:1.82] & & 0.58\ [0.52:0.64] &
 1.29\ [1.12:1.36] \\
OCC & Occitanie &  3.26\ [3.01:3.54] & 1.85 \ [1.56:2.66] & & 0.61\ [0.57:0.64] &
 0.49\ [0.33:0.63] \\
ARA & Auvergne-Rh\^one-Alpes&  2.96\ [2.68:3.31] & 4.52 \ [3.52:5.44] & & 0.68\ [0.65:0.71] &
 0.77\ [0.74:0.81] \\
PAC & Provence:C\^ote d'Azur &  2.82\ [2.47:3.14] & 2.79 \ [2.35:3.19] & & 0.78\ [0.75:0.81] &
 0.98\ [0.90:1.08] \\
COR & Corse &  5.47\ [4.87:5.88] & 1.37 \ [1.06:2.02] & & 0.01\ [0.00:0.04] &
1.26\ [0.93:2.01] \\
& {\bf Weighted Average} &  {\bf 3.43}\ [3.31:3.51] & {\bf 2.84} \ [2.65:3.10] & &
{\bf 0.65}\ [0.61:0.69] & {\bf 0.88}\ [0.86:0.93] \\
      \hline
    \end{tabular}
    }
  \end{center}
  \label{tab:R0}
\end{table}

\begin{figure}[htb]
  \centering
    \includegraphics[width=\hsize]{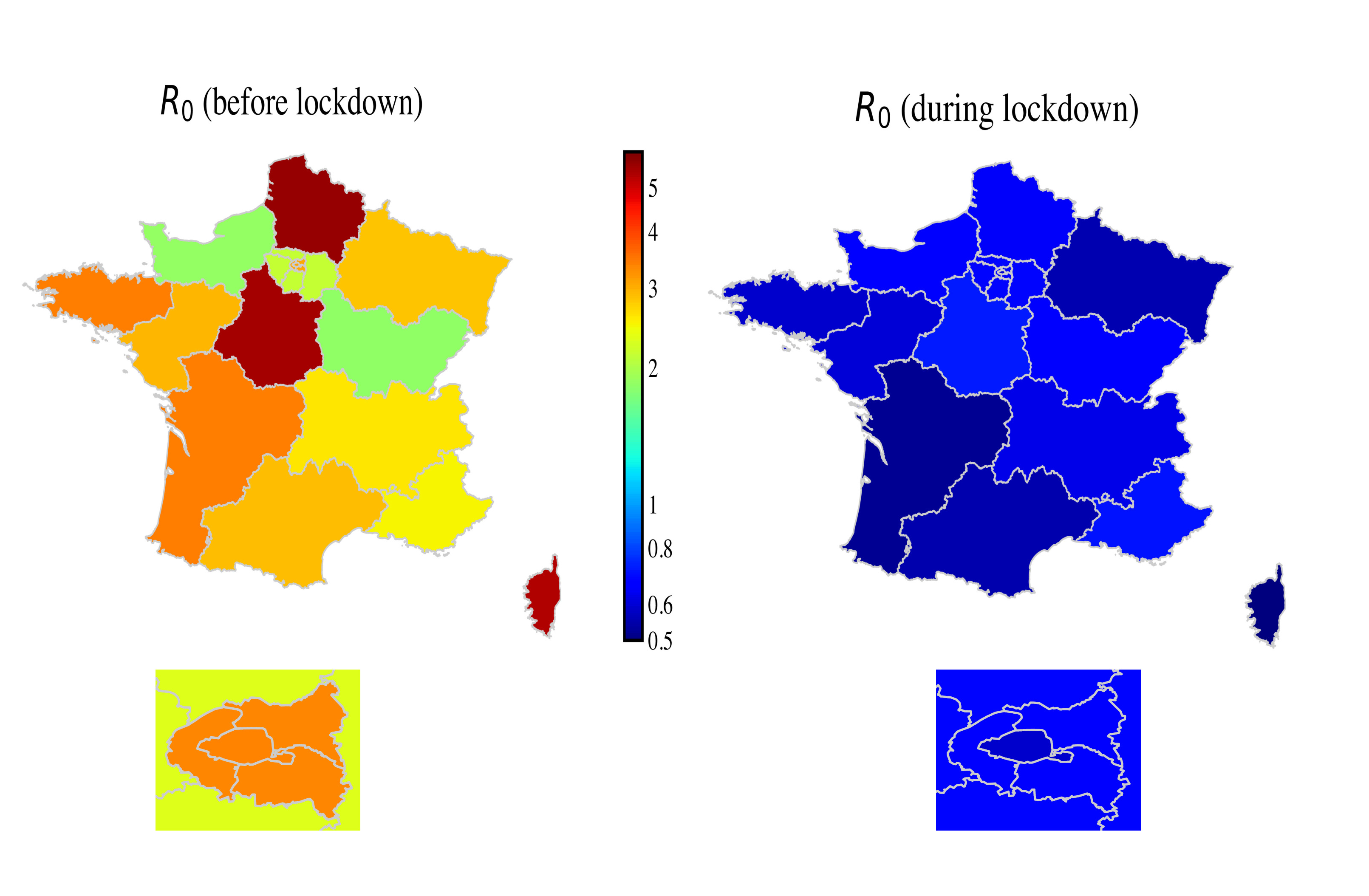} 
%%   \includegraphics[width=0.4\hsize,height=6cm]{Regions_Gen_m3_H1_w119_M11_s50000_t60_300_R0mapIni_simp.pdf}
%%   \includegraphics[width=0.35\hsize,height=6cm]{Regions_Gen_m3_H1_w119_M11_s50000_t60_300_R0mapConf_simp2.pdf}
%% \\
%%   \includegraphics[width=0.12\hsize,height=1.6cm]{Regions_Gen_m3_H1_w119_M11_s50000_t60_300_R0mapIni_IDFsmall.pdf}
%% \qquad\qquad\qquad\qquad\qquad\qquad\qquad\qquad
%% \includegraphics[width=0.12\hsize,height=1.6cm]{Regions_Gen_m3_H1_w119_M11_s50000_t60_300_R0mapConf_IDFsmall.pdf}
%% \qquad
  \caption{
Maps of basic reproduction factors in France (\emph{top}) and Paris area
(\emph{bottom}) before (\emph{left}) and during
(\emph{right}) lockdown for the SEAFHCDRO model.
  }
\label{fig:mapRegions}
\end{figure}

The \rz\ values are listed in Table~\ref{tab:R0}.
The population-weighted mean values are given at the bottom row of the Table.
The mean \rz\ values before lockdown are 3.43 [3.31 to 3.51] for SEAFHCDRO
and 2.84 [2.65 to 3.10] for SEAFHDRO (showing 90\% confidence intervals in brackets).
The corresponding values during lockdown are 0.65 [0.61 to 0.69] for
SEAFHCDRO and 0.88 [0.86 to 0.93] for SEAFHDRO.
Therefore the lockdown had a strong effect as it reduced \rz\ by  factor of
over 5 (SEAFHCDRO) or over 3 (SEAFHDRO).
This can graphically seen in Figure~\ref{fig:mapRegions}, which confirms the
overall success of the lockdown.

The city of Paris displays a high \rz\ before the lockdown,
which is expected given its very large population density.
Surprisingly, region Centre Val de Loire displays even higher \rz\ for both
models.
Region Grand Est, where the pandemic hit early and hard,
%% which suffered the highest rate of  hospital
%% visits and deaths, 
shows a low pre-lockdown \rz, probably because its
inhabitants self-confined well before the lockdown.
This is also true for Paris, whose inhabitants knew they were at great risk
because of the high population density, which may explain why its pre-lockdown
\rz\ is not even higher.

However, there are important differences in the marginal distributions of
\rz\ between the 2 models, as displayed in Figure~\ref{fig:margR0}.
Some regions are among the lowest in pre-lockdown \rz\ with SEAFHDRO, while
they are among the highest with SEAFHCDRO, e.g. Paris-Petite-Ceinture and
Hauts de France. With SEAFHDRO,
Pays de Loire and Centre Val de Loire are at respectively  the lower and
upper limits of our pre-lockdown
prior ($R_0=1$, see Table~\ref{tab:priors}), while they show normal values of \rz\ with SEAFHCDRO.
While the \rz\ values during the lockdown are all below unity for SEAFHCDRO,  a few
regions with the SEAFHDRO model show lockdown \rz\ values above unity:
Centre val de Loire, Bretagne, and especially
Nouvelle Aquitaine and Corse.

These issues suggests that SEAFHCDRO is a better model
compared to its SEAFHDRO counterpart. In what follows, I will therefore
concentrate on the SEAFHCDRO model.

\subsection{Past and future evolution}
%\subsubsection{France as a single region}
%% I begin with the case where we mix the data from the different \deps\ into a
%% single region. In other words, only single national values of $R_0$ are
%% derived.

\subsubsection{Evolution with no lifting of lockdown}

\begin{figure}[ht]
  \centering
  \includegraphics[width=0.44\hsize]{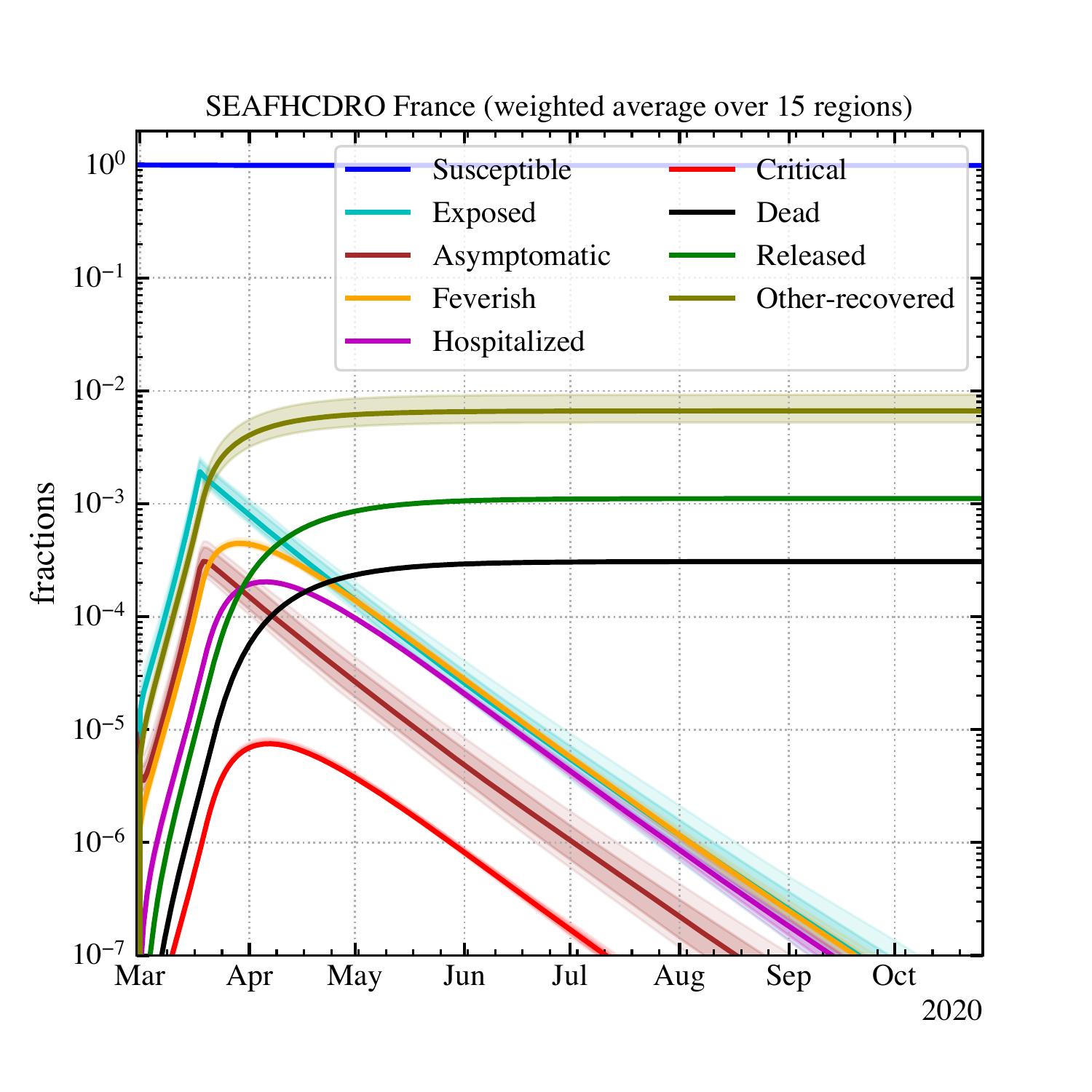} 
  \includegraphics[width=0.44\hsize]{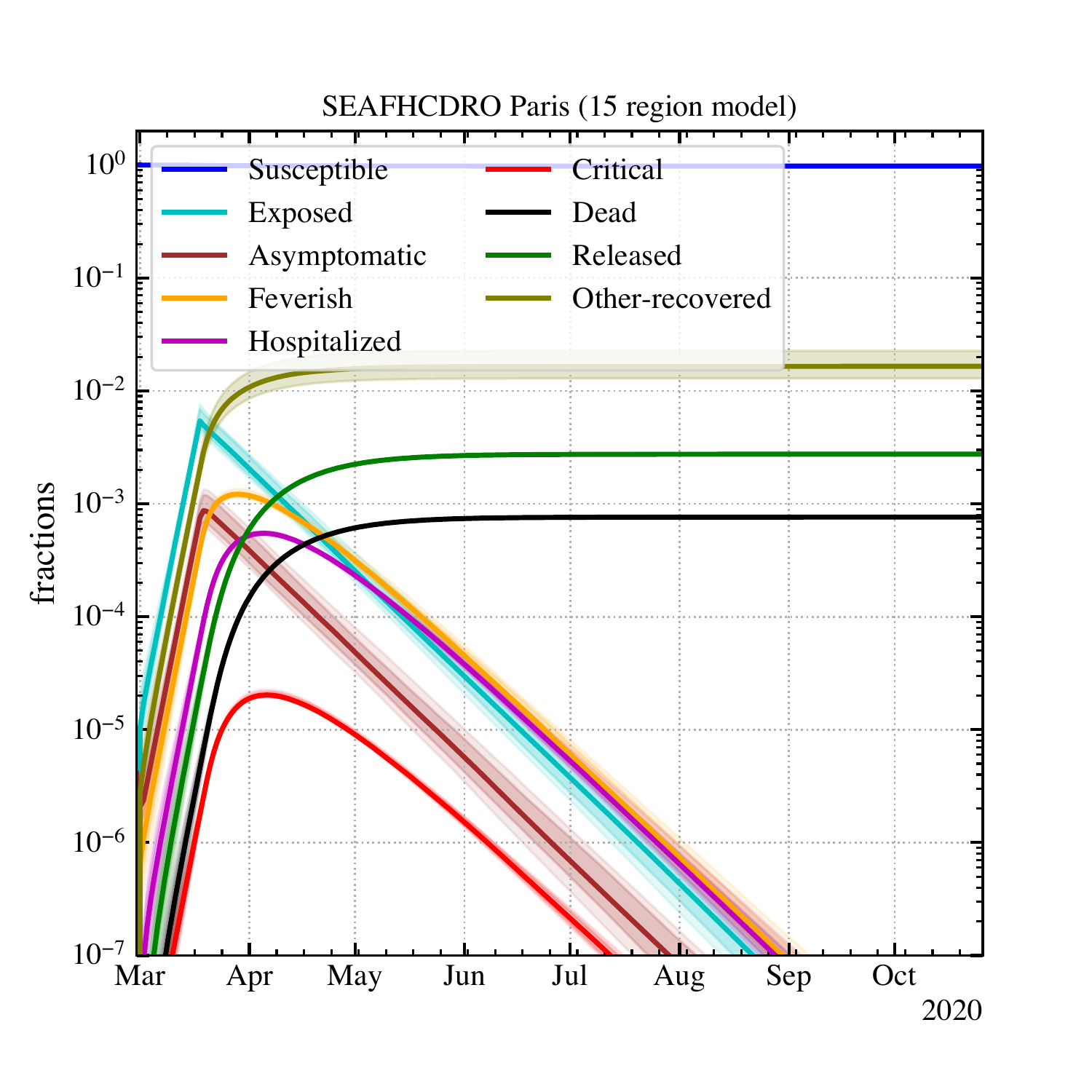}
  \includegraphics[width=0.44\hsize]{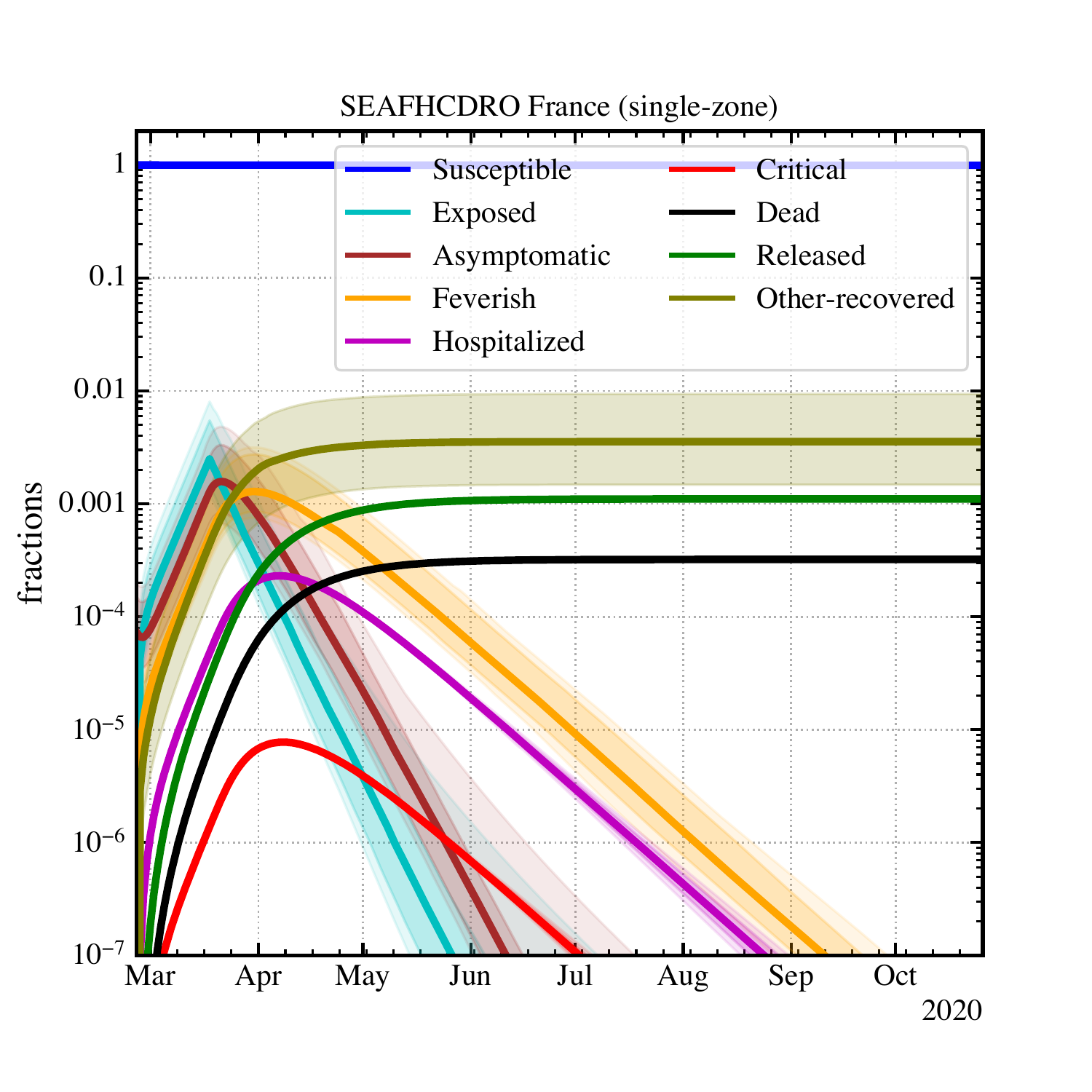}
  \includegraphics[width=0.44\hsize]{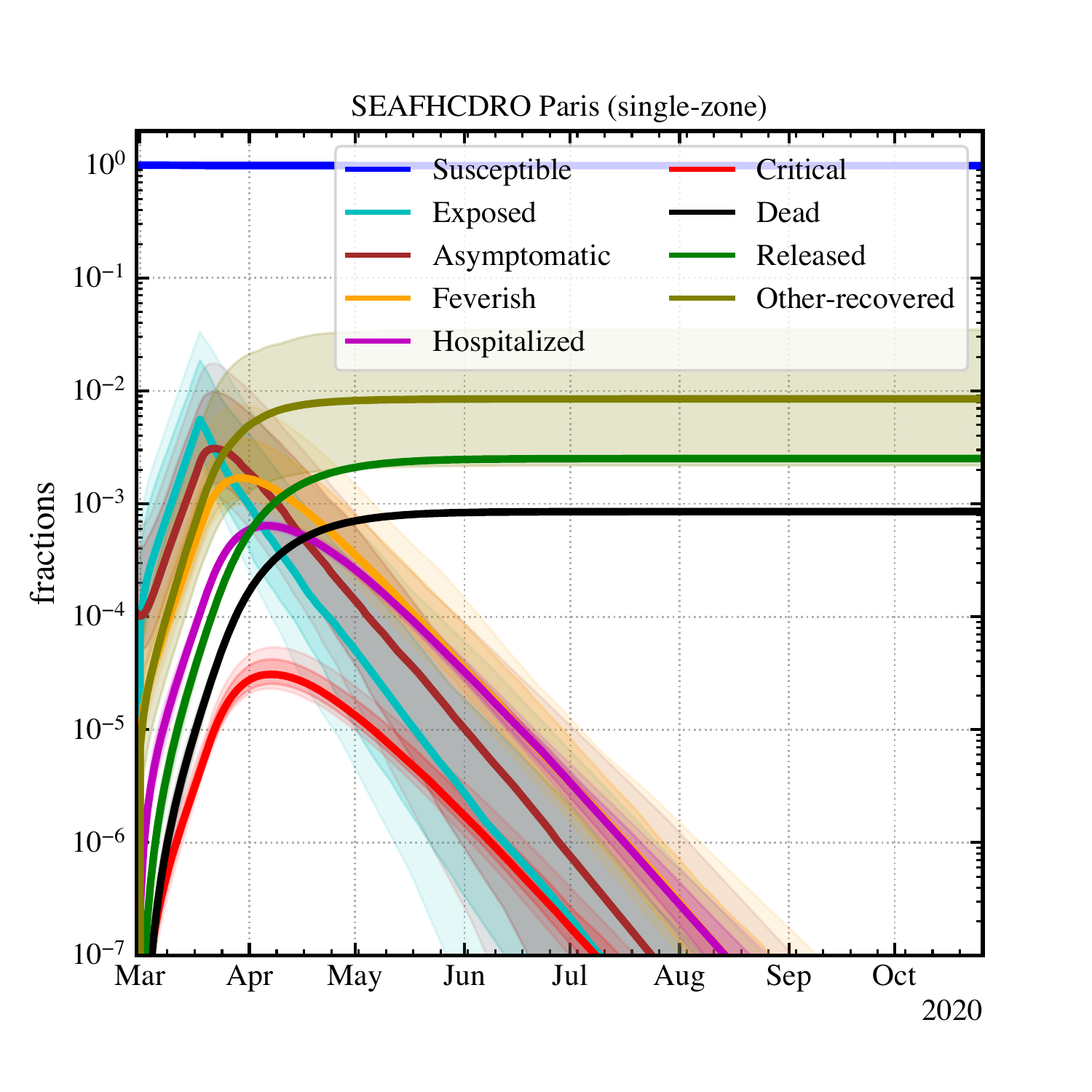} 
  \caption{
{\bf Top}: Evolution of the 9 SEAFHCDRO phases in the 15-region model for
France (weighted average, \emph{left}) and
Paris (\emph{right}).
The \emph{curves} are the mean post-burnin values, while the \emph{shaded} and
\emph{light-shaded areas} represent 16-85\% and 5-95\% confidence limits.
     {\bf Bottom}: Same for single-zone France ({\rm left}) and single-zone
     Paris (\emph{right}).
%     \textcolor{red}{Overplot true deaths as dashed curve in left figures.}
  }
\label{fig:evolMCMC}
\end{figure}

The evolution of the 9 SEAFHCDRO phases is shown in
Figure~\ref{fig:evolMCMC}.
%Both single-zone France and Paris
All four panels show interesting
features: the Exposed population has a sharp turn from exponential growth
before lockdown to exponential decrease once the lockdown has started.
The subsequent phases show increasingly more moderate and delayed reactions
to the lockdown.
Another interesting feature is that, SEAFHCDRO run on 15 regions allows fairly precise measures
of the different phases, including the early ones, which are naturally less
well constrained by the hospital data.
Despite having one-tenth of the hospital data of the country, the city of
Paris displays impressively precise estimates on the 9 phases. This is true
of all regions.

The bottom panels of Figure~\ref{fig:evolMCMC} show the same, but for
single-zone runs of France (left) and Paris (right).
Here, the Hospitalized, Critical, Dead and Released phases still show
low uncertainties, thanks to the available hospital data.
However, the uncertainties on the pre-Hospitalized phases
(Exposed, Asymptomatic, and Feverish) are now high.
In particular, while the Feverish (orange) curves are thin in the 15-region
(top) panels, they are very thick for the single-zone (bottom) ones.
This difference is related to the apparent degeneracy between terms $F$ and $\tfh/\fFH$
in equation~\ref{HdotSEAFHCDRO}. Without extra information, the degeneracy
causes a strong uncertainty on the Feverish population fraction, $F$. But
with the additional knowledge of the time-to-fraction ratios, such as
$\tfh/\fFH$ given by multi-zone runs, the degeneracy is lifted and the
uncertainty on $F$ is small.
This confirms the use of computer-expensive multiple-zone runs to better
constrain the time-to-branching-fraction ratios.

The top panels of Figure~\ref{fig:evolMCMC} indicate that
if the lockdown were not partially lifted on 11 May 2020, the number of Dead
would only very mildly rise, to fractions of 0.03\% for France as a whole
(i.e. 20 thousand deaths) and
0.08\% for Paris (1800 deaths). In fact on 11 May 2020, France had suffered
26\,000 deaths, and the underestimate from our model are within the
statistical uncertainties.
%The top panels of Figure~\ref{fig:evolMCMC} also indicate that
The huge majority of
the population remains Susceptible.

\begin{figure}[ht]
  \centering
  \includegraphics[width=0.49\hsize]{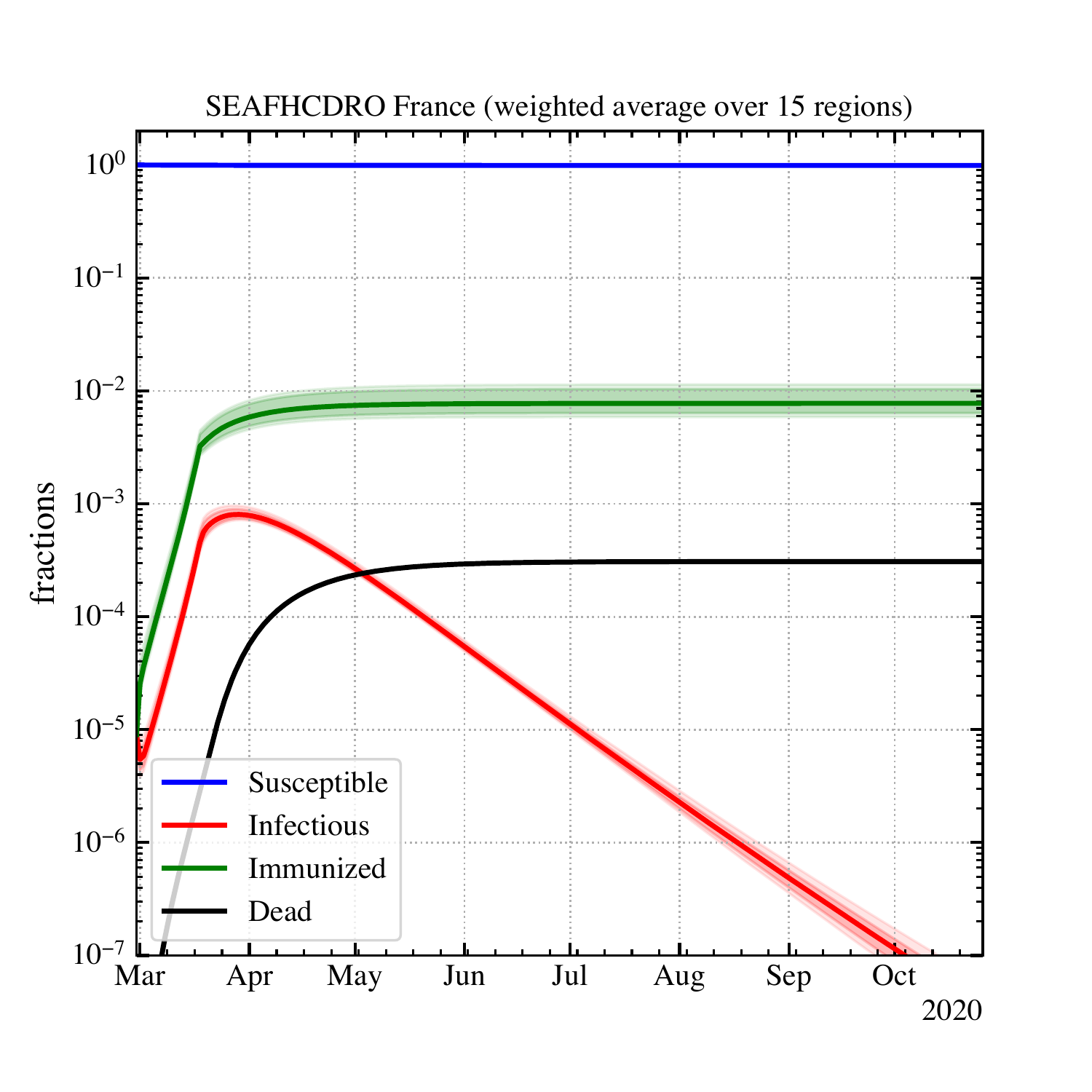}
  \includegraphics[width=0.49\hsize]{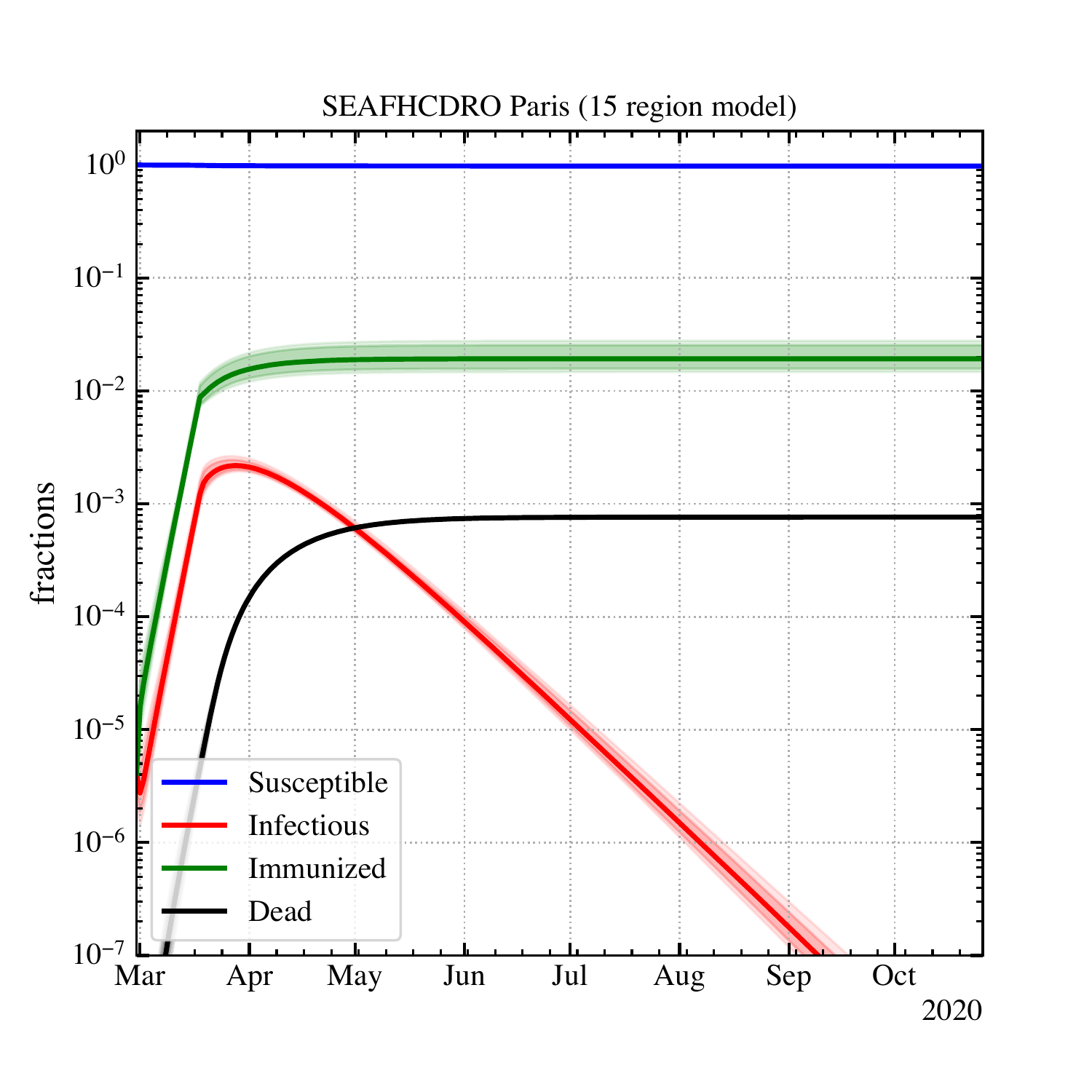}
  \caption{
Simplification of the 9 phases into 4 super-phases, for the 15-region model.
  }
\label{fig:4ways}
\end{figure}

Figure~\ref{fig:4ways} shows the evolution of the 4 \emph{super-phases}: Susceptible,
Infectious, Immunized, and Dead (see Fig.~\ref{fig:SEAFHCDRO}).
The total fraction of Infectious (i.e. Asymptomatic, Feverish, Hospitalized,
and Critical), which reached a maximum of several tenths of
one percent in the 2nd half of March, has decreased by a factor 10
(single-zone France) or 6 (Paris in 15-region analysis) by 11 May 2020. Thus,
on 11 May 2020, the total fraction of Infectious should be a few per ten
thousand.

\begin{figure}[ht]
  \centering
  \includegraphics[width=0.9\hsize,height=8cm]{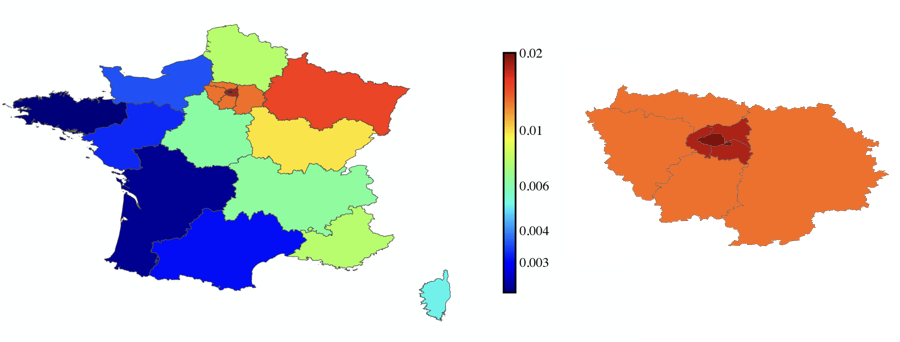}
  \caption{
Maps of fraction of Immunized in France on 11 May 2020, with SEAFHCDRO model
(median of marginal distributions).
  }
\label{fig:mapsImm}
\end{figure}

The fraction of Immunized (all but Susceptibles and Dead) is less than 1\% for all of France (weighted average
over the 15 regions), and 3\% for Paris, both at the 95\% confidence level. A
map of France of the fraction of Immunized on 11 May 2020 is displayed in
Figure~\ref{fig:mapsImm}. One notices that Paris and the Grand Est region
have the highest immunized fractions up to 2\%, followed by the inner ring around
Paris. The lowest fractions of Immunized all lie in the West of France, where
the fraction of Immunized is 10 times lower than in Paris.

These low fractions of Immunized means that France has not reached herd
immunity and is therefore sensitive to a second wave of the pandemic.
The fraction of Immunized is lower in the SEAFHDRO model (not shown in
Fig.~\ref{fig:4ways}), at less than 0.3\% on May 11,
rising to over 1\% in August, again insufficient for  herd immunity.

Figure~\ref{fig:4ways} allows one to estimate the 
Infectious Fatality Rate, IFR = Deaths/Immunized. After early July 2020, the
IFR stabilizes to values of 4\% (with a factor 30\% relative uncertainty) for 
both France and Paris with the 15-region
analysis. The values measured at 11 May 2020 are only slightly lower.
The SEAFHDRO model (not shown in Fig.~\ref{fig:4ways}) leads to a much higher
(but less plausible to this author) IFR of  12$\pm$2\% averaged over France.

These values are quite high, but France also currently has a very high
Case-Fatality-rate,
which has reached a plateau of 20\% since early May
2020.\footnote{France's CFR is second to Belgium in the Western
  world. But the estimates of CFR in France may suffer from discrepant
  sources, as deaths now include nursing homes (since early April), while
  cases may not.}

\subsubsection{How many deaths, had there been no lockdown nor social
  distancing?}

\begin{figure}[ht]
  \centering
  \includegraphics[width=0.49\hsize]{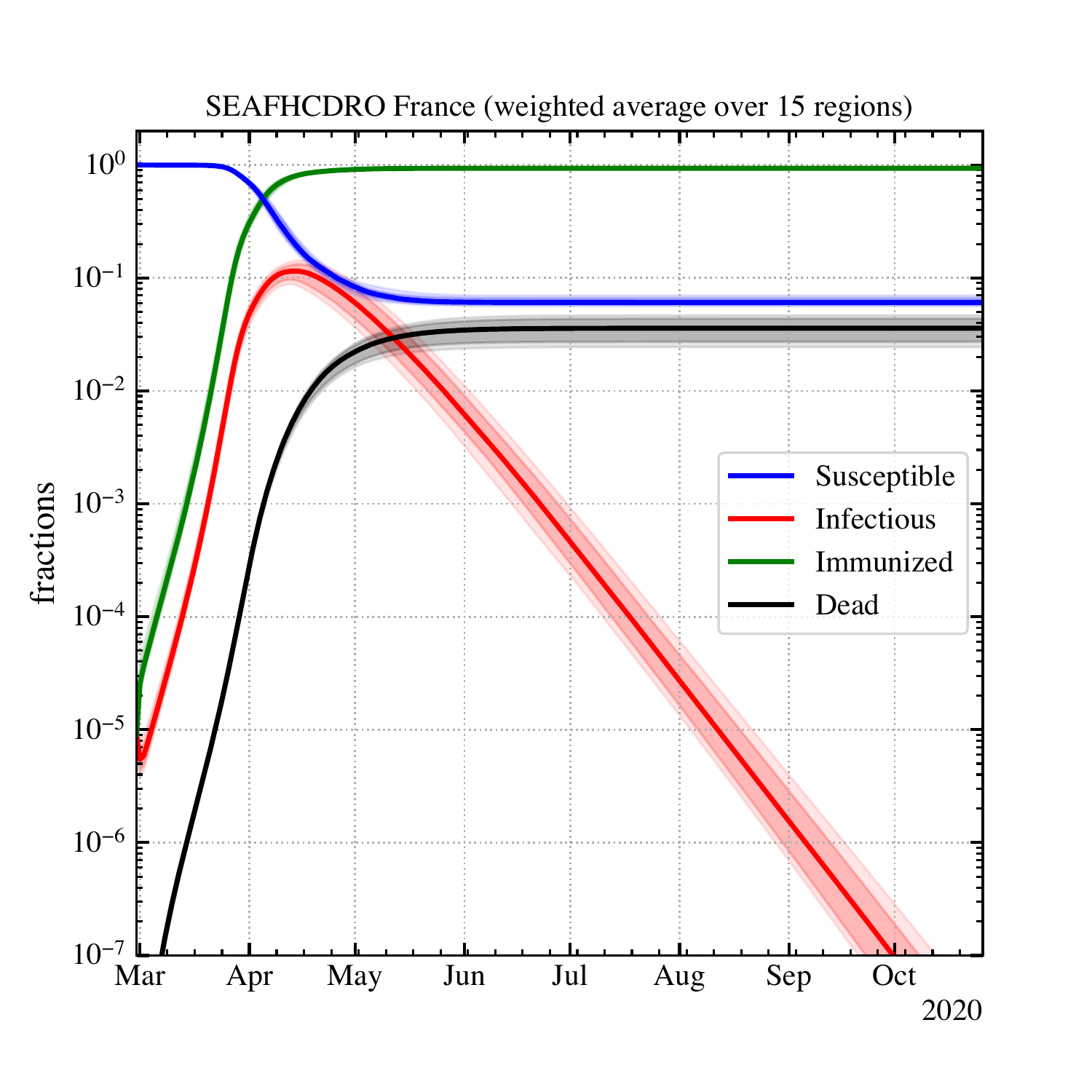} 
  \includegraphics[width=0.49\hsize]{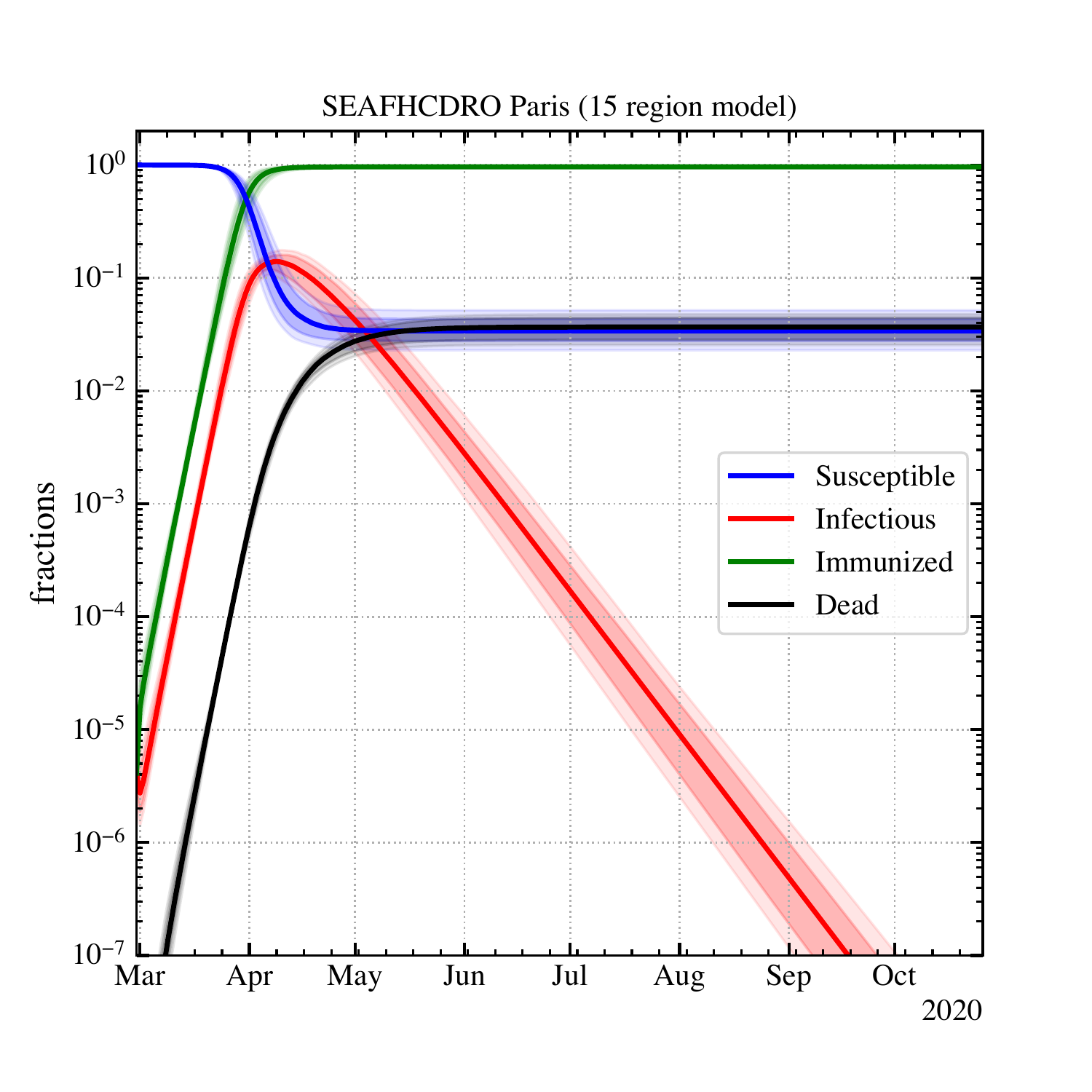}
  \caption{
          Prediction of evolution had there been:
      no lockdown on 17 March 2020 (\emph{left}) or a lockdown 10 days
      earlier (7 March 2020, \emph{right}), both with the
SEAFHCDRO model run with 15 regions, for France (weighted average).
  }
\label{fig:noConf}
\end{figure}

The modeling of the French hospital data presented here  allows to estimate the number of Dead in the absence of
lockdown and of social distancing.
It suffices to adopt the marginal distributions of the ratios of timescales
to branching fractions, as well as the marginal distributions of pre-lockdown
\rz\ factors for each region.

The left panel of Figure~\ref{fig:noConf} shows the evolution of the 4 super-phases in the
absence of any lockdown or social distancing with the SEAFHCDRO model.
The fraction of Dead would would have surpassed 1\% by mid-April. By early June 2020,
the fraction of Dead would have reached its plateau at 3.5$\pm$1\% (The
uncertainties are
90\% confidence) for France, and 4$\pm$1.5\% for Paris.
In other words, France would have suffered over 
2 million deaths without a lockdown!
Using instead the SEAFHDRO model (not shown in Fig.~\ref{fig:noConf}),
the fraction of Dead would have been 3 times
higher resulting in over 7 million deaths.
These are in fact lower limits, because the hospitals would been completely
saturated, leading to even more casualties.

\subsubsection{How many deaths would have been avoided with an earlier lockdown?}
One can similarly estimate the amount of deaths in France had the lockdown
been enforced 10 days earlier. Indeed, By 5 March 2020, France had suffered a
full week of daily doubling of cases and deaths, and the situation in the
Grand Est, in particular the city of Mulhouse was becoming out of
control. Assuming (probably optimistically) that it would take one day of decision-making and one day
for preparing the lockdown, the earliest possible starting date for the
national lockdown in France would have been on  7 March 2020. The right panel
of Figure~\ref{fig:noConf} shows the evolution of the super-phases with this
earlier lockdown. The fraction of deaths would have been reduced to less than
$1.4\times 10^{-5}$, i.e. less than 1000 deaths, instead of the 28\,000 on 22
May 2020.

\subsubsection{Scenarios for the evolution after the partial lifting of the
  lockdown}

\begin{figure}[ht]
  \centering
  \includegraphics[width=0.49\hsize]{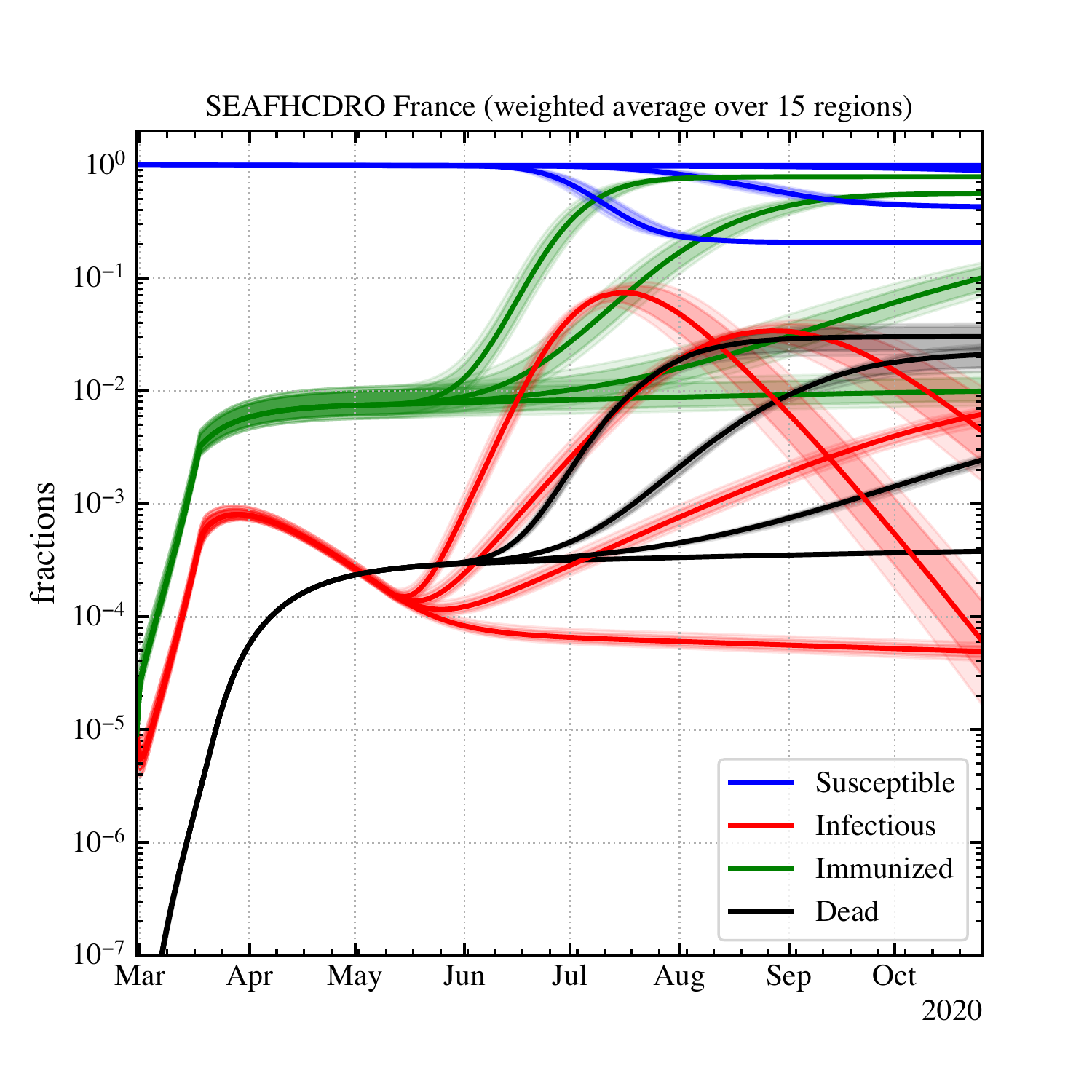}
\includegraphics[width=0.49\hsize]{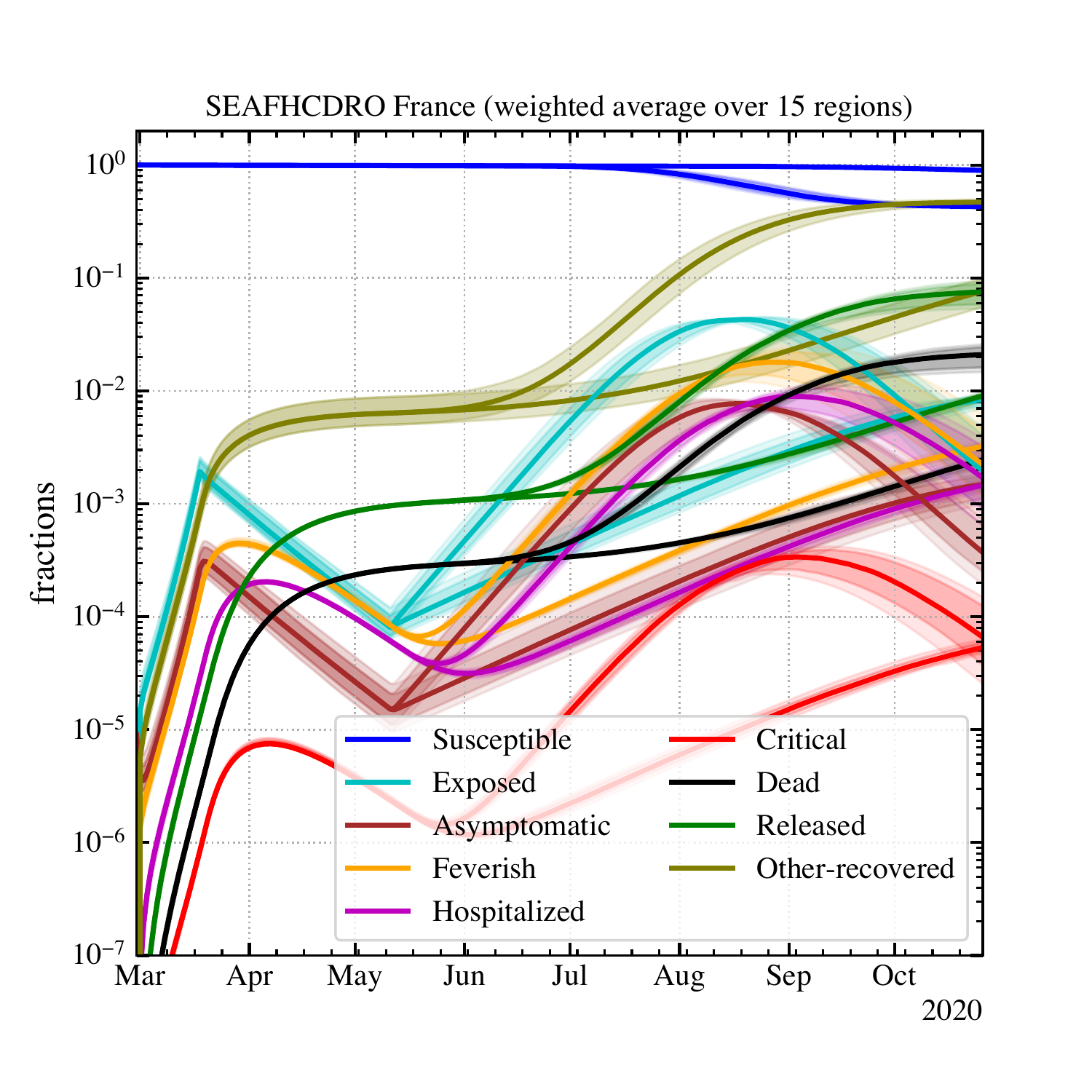}
  \caption{
Scenarios of evolution after 11 May 2020 using SEAFHCDRO for France (averaged over 15
regions).
The \emph{left panel} shows the 4 super-phases, with $R_0 = 1$, 1.2, 1.5, and
2, while the \emph{right panel} shows the 9 phases, with $R_0 = 1.2$ and
1.5.  In both panels \rz\ rises upwards for July 2020. Note that only blue
(Susceptible) and black (Dead) colors have the same meanings in the two
panels.
  }
\label{fig:PostMay11}
\end{figure}

The French authorities have partially lifted the lockdown on 11 May
2020. This should, by itself, create more encounters and thus raise
\rz. In the worst case, one would return to the pre-lockdown value of $R_) =
3.4$.
However, on the same date, face masks became widely available throughout
the country. If a fraction $f$ of the population wears a face mask, then
\rz\ is multiplied by the factor $(1-f)^2$. It suffices that $1-1/\sqrt{3.4}$
= 46\% of the
population wears a face mask to bring \rz\ below unity, and thus avoid a
second wave, even without any social 
distancing.  

Figure~\ref{fig:PostMay11} illustrates four scenarios for the evolution of the
pandemic in France after 11 May 2020, using 
pessimistic assumptions on \rz\ where face masks are
rarely used, with $R_0 = 1$, 1.2, 1.5, and 2.
Note that these values lie between the low lockdown and
the high pre-lockdown values of \rz.

The reader should first concentrate on the left panel of
Figure~\ref{fig:PostMay11}, showing the evolution of the four super-phases.
If $R_0 = 1$ after the partial lift of the lockdown, then the situation
remains stable, and the number of Dead barely increases (by 10\%).
If, instead, the public mingles too much, leading to a high value of $R_0 = 2$,
then the second wave will hit the country, with a peak of the Infectious
super-phase in July and a multiplication by 20 of the number of Dead,
reaching 3\% of the population of Paris and of France as a whole,
i.e., 2 million Dead!
Note that the evolution for France as a whole mimics that of each region,
i.e. Paris, because not only are the ratios of timescales to branching
fractions are the
same between regions, as before, but now the \rz\ factors are also assumed
the same.
If the public chooses an intermediate path, with $R_0 = 1.5$, then
the pandemic returns less dramatically than with $R_0 = 2$, and is delayed (as
predicted by many with the SIR model). Still, with \rz\ as high as 1.5 after
the partial lifting of the lockdown, the number of deaths would rise to
2\% of the population, i.e. over one million deaths!
Finally, if $R_0 = 1.2$, then the pandemic rise is very slow at the start,
with the number of Dead doubling only in August. But without stronger
containment, this scenario would also lead to 0.25\% of deaths in France,
i.e. over 150 thousand.
In other words, the time of doubling the total number of deaths in France on
11 May (26 thousand) is 1, 2, or 3 months if $R_0 = 2$, 1.5, or 1.2.
The evolution of the pandemic is therefore highly sensitive to the value of
\rz, with quiet evolution for $R_0 = 1$, manageable evolution for $R_0 = 1.2$,
and increasingly fast second waves for $R_0 = 1.5$ and $R_0 = 2$.

It is useful to compare the evolution of Dead and of Infectious people. The
dates when the number of Infectious will multiply by 5 compared to May 11 are
May 25, June 10, and July 22, for $R_0 = 1.2$, 1.5, and 2, respectively.
Similarly, the dates when the total number of Deaths in France will be
multiplied by 5 (to reach 130\,000) are June 28, July 31, and October 22, for
$R_0 = 1.2$, 1.5, and 2, respectively. This means that the delay between
Infectious and Dead super-phases is
34, 51, or 90 days, for $R_0 = 1.2$, 1.5, and 2, respectively.

The right panel of Figure~\ref{fig:PostMay11}, % admittedly very crowded,
allows to obtain delays of the
increase in Deaths with corresponding increases in earlier
phases.
The dates when the number of Hospitalized multiply by 5 since May 11, are
June 12, and July 24, for $R_0 = 1.2$ and 1.5, respectively,
i.e. very close to the dates of the Infectious in general. This is the
consequence of the Feverish and Hospitalized dominating the Infectious
super-phase on May 11, in contrast to the first-wave exponential growth when
the Asymptomatic phase dominated the Infectious super-phase (see also
Fig.~\ref{fig:evolMCMC}). 
This is good news for the French authorities, as they can use hospital data
to monitor the rise of deaths with over a month between the quintupling of
hospitalizations and the quintupling of deaths, as long as $R_0 \leq 2$.

\section{Conclusions and Discussion}

This work applies several  epidemiological models to fit the French
hospital data from March 19 to 4 May 2020. The evolution of
hospitalizations and deaths after the national lockdown  of 17 March was too
slow to be compatible with the SIR model (Fig.~\ref{fig:fIoft}).
While the SEIHCDRO model (Fig.~\ref{fig:SEIHCDRO}) provided an adequate fit
to the hospital data, these data were better fit with the
SEAFHCDRO model (Fig.~\ref{fig:SEAFHCDRO}),
which splits the Infectious phase between one (Asymptomatic) that
effectively
contaminates Susceptibles and the following one (Feverish) whose members are
too ill to leave their dwelling and contaminate Susceptibles.
This better fit for SEAFHCDRO has higher Bayesian evidence, even after allowing for the extra
two free parameters.

However, neither SEIHCDRO nor SEAFHCDRO can reproduce the rapid decrease in
daily arrivals of Critical people at hospitals, which is more rapid than
the corresponding decreases since lockdown in the daily Hospital arrivals, as
well as daily Releases and Deaths (Figs.~\ref{fig:GoFFrance} and \ref{fig:GoFRegionsParis}).
I then introduced the SEAFHDRO model (Fig.~\ref{fig:SEAFHDRO})
that merges the Critical and
Hospitalized phases, but this model produced unsatisfactory results:
1) it fits less well the decrease in the daily Deaths compared to
the SEAFHCDRO model;
2) it produces $R_0 > 1$ during lockdown for some regions.
%for the lockdown phase. 
%% The conclusions of this study on the future evolution of the pandemic
%% vary fairly little whether one adopts one or the other model.

This study produces the first analysis of ratios of timescales to branching
fractions without resorting to the results of viral (PCR) testing.
Such testing provides constraints on the timescales and branching fractions
for the evolutionary paths
$H\to C$, $H\to R$, $C\to D$, and $C\to R$, but not for previous paths, for
which random testing is required.
The results presented here indicate that the incubation phase lasts less than
5 days (Fig.~\ref{fig:margRatios}), in agreement with many studies based on
testing (e.g. \citealp{Lauer+20}).
Moreover, this study produces the first geographic analysis of the
\rz\ factors, both before lockdown and during the lockdown.
It indicates some regional differences, with Hauts-de-France and Paris
showing higher \rz\ before the lockdown, with a homogenization during the
lockdown.
Overall,
%% in agreement with previous studies
%% \citep{Roques+20,Unlu+20,DiDomenico+20,Salje+20},
the SEAFHCDRO model leads
to $R_0 = 3.4\pm0.1$ before and $R_0 = 0.65\pm0.02$ during the lockdown
(with 90\% confidence limits, see Table~\ref{tab:R0}),
with some regional variations (Figs.~\ref{fig:margR0}
and
\ref{fig:mapRegions}).
Therefore the national lockdown contributed to decreasing the basic reproduction number by a factor of
roughly 5.
The pre-lockdown value above is in good agreement
with the values previously reported for France: 
$R_0 = 3.2\pm0.1$ \citep{Roques+20},
$R_0 = 3.56$ (\citealp{Unlu+20}, without uncertainties),
$R_0 = 3.0\pm0.2$ \citep{DiDomenico+20}, and
$R_0 = 3.3\pm0.1$ and $3.4\pm0.1$ for the two models of \citep{Salje+20}.
Similarly, the lockdown value of \rz\  during lockdown is in
good agreement with previously
reported values:
$R_0 = 0.74$ (\citealp{Unlu+20}, with no uncertainties),
$R_0 = 0.68\pm0.06$ \citep{DiDomenico+20},
$R_0 = 0.47\pm0.03$ \citep{Roques+20b},
and
$R_0 = 0.52\pm0.03$  for both models of \citep{Salje+20}.

%% The Infection-Fatality Rate can be estimated with the SEAFHCDRO model, which
%% leads to IFR = 4$\pm$1\%, which is 8 times higher 

The SEAFHCDRO model also predicts the evolution of the 9 phases, and was also
run in scenarios of no lockdown and partial lifting of the lockdown.
The use of multiple-zone fits allows to accurately predict the early phases:
Exposed, Asymptomatic and Feverish phases, which the single-zone models
cannot do (Fig.~\ref{fig:evolMCMC}).
On 11 May 2020, the Feverish phase dominates the Infectious super-phase,
followed closely by the Hospitalized, which both vastly outnumber the
Asymptomatics who dominated
the Infectious super-phase during the early pre-lockdown exponential growth
(Fig.~\ref{fig:evolMCMC}).

The time variations of the fractions (hence populations) of the different
phases allow the measure of the fraction of 
Immunized, which is less than 1\% throughout France and 3\% in Paris (both
with 95\% confidence) with both the SEIHCDRO and SEAFHCDRO models (and 10
times less with SEAFHDRO). The first fraction is 4 times lower than the
fraction of $4\pm0.1\%$ found by \citet{Roques+20b}.

The IFR of France is estimated at 4$\pm$1\%,
i.e. 5 to 6 times
higher than the previous estimates of \citet{Roques+20} and \citet{Salje+20}.
This difference may arise from incorrect assumptions on timescales and fractions by
those two teams, or conversely by the lack of sufficiently strong
priors on timescale to fraction
ratios in the present analysis.

One should not be tempted by the argument that some fraction of people
with high fever may still go out and contaminate Susceptibles. Indeed,
the
separation of Infectious into Asymptomatic and Feverish is precisely based on
the assumption that what are called Feverish do not contaminate Susceptibles.
In other words, by Feverish, one may think of people with very high fever,
who will not only remain at home, but self-isolate from the other people in
their dwelling.

Had there been no lockdown, nor social distancing, the COVID-19 pandemic
would have resulted in over 2 million deaths in France
(Fig.~\ref{fig:noConf}).
Conversely, had the lockdown been enforced 10 days earlier, the number of
deaths would have been 30 times lower.
After May 11, the evolution will naturally depend on the value of the
\rz\ factor, which will be set by how well will people respect social
distancing and on the fraction of people wearing facial masks. If nearly half
of the population is masked, then \rz\ should stay below unity and the
pandemic will keep decreasing. On the other hand, if fewer than 46\% of the
population wear face masks, and assuming little social distancing, the
pandemic will suffer a second wave (see
Fig.~\ref{fig:PostMay11}).
A value of \rz\ as `low' as 1.5 would produce disastrous consequences by the
end of the Summer, if left unimpeded. But if France achieves $R_0 = 1.2$, it
will have much more time to react to rises in hospitalizations.

There are several caveats to these results (the first few affecting many other
similar studies).
1) Many deaths occur outside of hospitals, hence the dire deaths estimates of
the models presented here are probably underestimates.
2) The models neglect Deaths of COVID-19 Hospitalized patients that were not
sent to Critical care.
3) The models assume non-evolving ratios of timescales over branching
fractions.
However, the arrivals of Critical patients decreases faster than the
  SEAFHCDRO model  predicts and the daily Deaths decreases faster than the
  models at late times, both suggesting that with time doctors have devised
  better protocols in handling
  COVID-19 patients without the need to send them to Critical care, and better
  treatment of the Critical care patients.
4) The analysis fails to follow the age distribution of the hospital data,
whereas deaths by COVID-19 increase exponentially with age (as do regular
deaths).

Nevertheless, it is hoped that this sort of analysis will present an example
for future studies. It could be enhanced with better testing data, in
particular regular random testing of a small fraction of the population, as
well as other indicators of SARS-2 (e.g. the analysis of SARS-2 viruses in
sewage \citealt{Wurtzer+20}). 

\section*{Acknowledgements}

I am grateful to Avishai Dekel who introduced me to SIR and snowed me with
his very pretty mathematical developments, Trevor Ponman for useful
discussions with provocative questions, as well as Lionel Roques,  Vittoria
Colizza, and Andr\'e Klarsfeld for useful discussions.
I also thank A. Klarsfeld, Eliott Mamon, Gabrielle Mamon and Nick Kaiser for useful
references.
This work has made use of the Horizon Cluster hosted by Institut d'Astrophysique de Paris.
I thank Christophe Pichon for access to this machine
and St\'ephane Rouberol for running smoothly this cluster for me (including
replying to my message on evenings and weekends).
I also am very grateful to 
Boris Dintrans for quasi-immediate supply of substantial computing power on the
OCCIGEN super-computer operated by CINES, which I unfortunately was not able
to process for this study.
%% Special thanks go to
%% St\'ephane Rouberol (Horizon) as well as
%% Eric Boyer and G\'erard Gil (CINES)
%% for
%% useful help (including late evenings and weekends).
Finally thanks to Eduardo Vitral and Matthieu Tricottet for help with Python
programming and Nikos Prantzos and David Valls-Gabaud for spotting an
arXiv-induced typo in eq.~(1) of an earlier version.
I also thank the makers of public software {\sc Python} with many packages,
especially
{\sc Scipy},
{\sc Numpy} \citep{vanderWalt11},
{\sc pandas} \citep{McKinney10},
{\sc emcee} \citep{Goodman&Weare10},
{\sc Matplotlib} \citep{Hunter07}, as well as Pierre Raybaut for developing
the {\sc Spyder} Integrated Development Environment.

\bibliography{covid}
\end{document}